\newcommand{\lya}{\ifmmode { Ly}\alpha \else Ly$\alpha$\fi}
\documentclass[longauth]{aa} 
\usepackage{graphicx}
\usepackage{amssymb}
\usepackage{cuted}
\usepackage{longtable}
\usepackage{booktabs}
\usepackage{caption}
\usepackage{txfonts}
\usepackage{longtable}
\usepackage{xcolor}
\usepackage{hyperref}
\hypersetup{colorlinks, 
 linkcolor = blue,
 urlcolor = blue,
 citecolor = blue,
 anchorcolor = blue}
\begin{document} 

 \title{Little impact of mergers and galaxy morphology on the production and escape of ionizing photons in the early Universe}
 \titlerunning{Impact of mergers and morphology on ionizing photon escape in the early Universe}
 
 \author{S. Mascia\fnmsep\thanks{E-mail: sara.mascia@ista.ac.at}
 \inst{1,2}
 \and
 L. Pentericci 
 \inst{1}
 \and 
 M. Llerena
 \inst{1}
 \and
 A. Calabrò
 \inst{1}
 \and 
 J. Matthee
 \inst{2}
 \and
 S. Flury
 \inst{3}
 \and
 F. Pacucci
 \inst{4}
 \and 
 A. Jaskot
 \inst{5}
 \and
 R. O. Amorín 
 \inst{6}
 \and
 R. Bhatawdekar
 \inst{7}
 \and
 M. Castellano
 \inst{1}
 \and
 N. Cleri
 \inst{8,9,10}
 \and
 L. Costantin
 \inst{11}
 \and
 K. Davis
 \inst{12}
 \and
 C. Di Cesare
 \inst{2}
 \and
 M. Dickinson
 \inst{14}
 \and
 A. Fontana
 \inst{1}
 \and
 Y. Guo
 \inst{14}
 \and
 M. Giavalisco
 \inst{13}
 \and
 B. W. Holwerda
 \inst{15}
 \and
 W. Hu
 \inst{16,17}
 \and
 M. Huertas-Company
 \inst{18}
 \and
 Intae Jung
 \inst{19}
 \and
 J. Kartaltepe
 \inst{20}
 \and
 D. Kashino
 \inst{21}
 \and
 Anton M. Koekemoer
 \inst{20}
 \and
 R. A. Lucas
 \inst{20}
 \and
 J. Lotz
 \inst{20}
 \and
 L. Napolitano
 \inst{1}
 \and
 S. Jogee
 \inst{24}
 \and
 S. Wilkins
 \inst{22,23}
 }
 
 \institute{
 INAF – Osservatorio Astronomico di Roma, via Frascati 33, 00078, Monteporzio Catone, Italy 
 \and 
 Institute of Science and Technology Austria (ISTA), Am Campus 1, A-3400 Klosterneuburg, Austria 
 \and 
 Institute for Astronomy, University of Edinburgh, Royal Observatory, Edinburgh, EH9 3HJ, UK
 \and 
 Center for Astrophysics, Harvard \& Smithsonian, Cambridge, MA 02138, USA
 \and 
 Department of Physics and Astronomy, Williams College, Williamstown, MA 01267, USA
 \and 
 Instituto de Astrof\'{i}sica de Andaluc\'{i}a (CSIC), Apartado 3004, 18080 Granada, Spain
 \and 
 European Space Agency (ESA), European Space Astronomy Centre (ESAC), Camino Bajo del Castillo s/n, 28692 Villanueva de la Cañada, Madrid, Spain
 \and 
 Department of Astronomy and Astrophysics, The Pennsylvania State University, University Park, PA 16802, USA
 \and 
 Institute for Computational and Data Sciences, The Pennsylvania State University, University Park, PA 16802, USA
 \and 
 Institute for Gravitation and the Cosmos, The Pennsylvania State University, University Park, PA 16802, USA
 \and 
 Centro de Astrobiología, Consejo Superior de Investigaciones Científicas, Madrid, Spain
 \and 
 Department of Physics, 196A Auditorium Road, Unit 3046, University of Connecticut, Storrs, CT 06269, USA
 \and 
 University of Massachusetts Amherst, 710 North Pleasant Street, Amherst, MA 01003-9305, USA
 \and 
NSF’s National Optical-Infrared Astronomy Research Laboratory, 950 N. Cherry Ave., Tucson, AZ 85719, USA
 \and 
 University of Louisville, Department of Physics and Astronomy, 102 Natural Science Building, Louisville KY 40292, USA
 \and 
 Department of Physics and Astronomy, Texas A\&M University, College Station, TX 77843-4242, USA
 \and 
 George P. and Cynthia Woods Mitchell Institute for Fundamental Physics and Astronomy, Texas A\&M University, College Station, TX
 \and 
 Instituto de Astrofísica de Canarias - C/Vía Láctea s/n-La Laguna - Spain
 \and 
 Space Telescope Science Institute, 3700 San Martin Drive, Baltimore, MD 21218, USA
 \and 
 Laboratory for Multiwavelength Astrophysics, School of Physics and Astronomy, Rochester Institute of Technology, 84 Lomb Memorial Drive, Rochester, NY 14623, USA
 \and 
National Astronomical Observatory of Japan, 2-21-1 Osawa, Mitaka, Tokyo 181-8588, Japan
 \and 
 Astronomy Centre, University of Sussex, Falmer, Brighton BN1 9QH, UK
 \and 
 Institute of Space Sciences and Astronomy, University of Malta, Msida MSD 2080, Malta
 \and 
 The University of Texas, Austin, TX, USA
 }

 \date{Received XXX; accepted XXX}

 
 \abstract
 {Compact, star-forming galaxies with high star formation rate surface densities ($\Sigma_{\text{SFR}}$) are often efficient Lyman continuum (LyC) emitters at $z\leq 4.5$, likely as intense stellar feedback creates low-density channels that allow photons to escape. Irregular or disturbed morphologies, such as those resulting from mergers, can also facilitate LyC escape by creating anisotropic gas distributions. We investigate the influence of galaxy morphology on LyC production and escape at redshifts $5 \leq z \leq 7$ using observations from various \textit{James Webb Space Telescope} (JWST) surveys. Our sample consists of 436 sources, which are predominantly low-mass ($\sim 10^{8.15} M_\odot$), star-forming galaxies with ionizing photon efficiency ($\xi_{\rm ion}$) values consistent with canonical expectations.
 Since direct measurements of $f_{\rm esc}$ are not possible during the Epoch of Reionization (EoR), we predict $f_{\rm esc}$ for high-redshift galaxies by applying survival analysis to a subsample of LyC emitters from the Low-Redshift Lyman Continuum Survey (LzLCS), selected to be direct analogs of reionization-era galaxies. We find that these galaxies exhibit on average modest predicted escape fractions ($\sim 0.04$). Additionally, we assess the correlation between morphological features and LyC emission. Our findings indicate that neither $\xi_{\rm ion}$ nor the predicted $f_{\rm esc}$ values show a significant correlation with the presence of merger signatures. This suggests that in low-mass galaxies at $z \geq 5$, strong morphological disturbances are not the primary mechanism driving LyC emission and leakage. Instead, compactness and star formation activity likely play a more pivotal role in regulating LyC escape.}
 \maketitle
\section{Introduction}

The Epoch of Reionization (EoR) marks a crucial phase in the Universe's history, beginning with the formation of the first stars at $z \sim 20-30$ \citep[e.g.,][]{Barkana2001} and concluding around $z \sim 5.5$, when reionization was complete \citep[e.g.,][]{Bosman2022}.
During this period, the first luminous sources, such as active galactic nuclei (AGNs) and massive stars in the early galaxies, emitted ionizing photons that reionized the neutral hydrogen in the intergalactic medium (IGM). While AGNs have been proposed as potential contributors to reionization \citep[e.g.,][]{Madau15, Finkelstein2019, Madau2024, Dayal2024}, the consensus is that young, hot stars in star-forming galaxies were the primary drivers \citep{Robertson2013, atek2023, Asthana2024}. These stars produced ultraviolet (UV) radiation, specifically in the Lyman continuum (LyC, $\lambda$ < 912 \AA), which ionized the surrounding hydrogen and played a key role in transforming the early Universe.

The ability of these galaxies to provide the necessary number of ionizing photons to reionize the Universe depends on the escape fraction ($f_{\rm esc}$) of the LyC photons, the percentage of LyC photons that successfully escape the galaxies' interstellar medium (ISM) and circumgalactic medium (CGM) to reach the IGM. An average $f_{\rm esc}$ of around 0.1 across all galaxies is estimated to be necessary to match the reionization history, as predicted by independent constraints including cosmic microwave background (CMB) data \citep{Planck2020} and high-redshift galaxy luminosity functions \citep[e.g.,][]{Robertson2015, Yung2020a, Yung2020b, Finkelstein2019}. However, directly measuring $f_{\rm esc}$ at $z \geq 4.5$, where reionization occurred, is impossible due to the increasing IGM opacity at that epoch \citep{inoue2014}.

One avenue to address this challenge is to study low-redshift galaxies with measurable LyC leakage, to search for properties that are analogous to those of galaxies in the EoR.
These lower-redshift galaxies, known as Lyman continuum emitters (LCEs), allow astronomers to measure LyC escape fractions and explore the conditions in the ISM that facilitate LyC photon escape \citep{Izotov16, Izotov_2020, Verhamme2017, Schaerer2022, Saxena2022, Flury2022, Flury2022b, Chisholm2022, Saldana-Lopez2022}. These measurements are then used to infer the average $f_{\rm esc}$ of the cosmic reionizers 
\citep[e.g.,][]{Jung2023, Mascia2023_Glass, Mascia2024_CEERS, Roy2023, Saxena2023b, Li2024}, consistently finding an average $f_{\rm esc}$ lower than 0.1.

Among the many programs analyzing the properties of low-redshift LyC leakers, the most comprehensive is the LzLCS survey (GO 15626, PI Jaskot) \citep{Flury2022, Flury2022b, Saldana-Lopez2022}. The LzLCS sample was augmented with archival COS observations from \cite{Izotov2016a, Izotov2018a, Izotov2018b, Izotov2021, Wang2019}. The combined dataset, referred to as the LzLCS+ sample, includes 88 low-redshift galaxies with either detections or stringent limits on LyC emission. The survey tested various indirect diagnostics to understand their correlation with LyC emission and found that indicators based on \lya\ emission are particularly effective: LyC emitters tend to have a high \lya\ escape fraction and a small velocity peak separation. 
Additionally, properties such as high [\textrm{O}\textsc{iii}]$\lambda\lambda4960, 5008$/[\textrm{O}\textsc{ii}] $\lambda$3727 (O32) flux ratio, high specific star formation rate surface density ($\Sigma_{\text{sSFR}}$), and a blue UV slope $\beta$ \citep{Chisholm2022} also correlate with a high LyC escape \citep{Flury2022}.

Although many properties correlate with $f_{\rm esc}$, such relations are extremely scattered: for this reason, several authors have refined methods to predict LyC emission using a combination of more properties, the so-called multivariate diagnostics. Initial studies \citep{Mascia2023_Glass, Mascia2024_CEERS, Lin2023} employed traditional regression methods, However, these models often struggle with censored data -- datasets that include upper limits rather than exact measurements. Additionally, these methods assume a linear relationship between the variables and $f_{\rm esc}$, a condition that may not always be met. Recently, \cite{Jaskot24a, Jaskot24b} proposed survival analysis techniques, which are better suited to the nature of the LzLC+ dataset, with its broad range of $f_{\rm esc}$ values and numerous upper limits. Ignoring these upper limits could lead to biased predictions and inaccurate insights into what differentiates galaxies that emit LyC radiation from those that do not \citep[e.g.,][]{Isobe1986}. 
One of the most promising models 
is the TopThree \citep{Jaskot24b}, which incorporates three key predictors: the star formation rate surface density ($\Sigma_{\text{SFR}} = \text{SFR}/(2\pi r_e^2)$, where $r_e$ is the UV half-light radius), the O32 ratio, and the UV slope ($\beta$). 
It was already noted that a high $\Sigma_{\text{SFR}}$ correlates with efficient LyC photon escape \citep[e.g.,][]{Heckman2001, Naidu2020, Flury2022}, particularly in compact galaxies where intense star formation feedback creates low-density channels in the ISM. These channels facilitate LyC photon escape, resulting in a higher $f_{\rm esc}$. 
For instance, J1316+2614, a galaxy at $z = 3.613$, is noted as one of the brightest LyC emitters known, with an escape fraction of approximately 0.9. This galaxy’s compact morphology, characterized by a half-light radius of $r_{e} \sim 0.22$ kpc and an extremely high $\log_{10} \Sigma_{\text{SFR}} \approx 3.47 \text{M}_\odot \text{yr}^{-1} \text{kpc}^2$, suggests that intense star formation efficiency plays a critical role in driving LyC escape \citep{Marques-Chaves2022, Marques-Chaves2024}.

Although compactness and high $\Sigma_{\text{SFR}}$ are common in many LCEs, this is not universally true. Some LCEs exhibit more extended or irregular structures, likely shaped by processes such as galaxy mergers or interactions \citep[e.g.,][]{Bergvall2013}. A notable merging system with detected LyC flux is Haro 11 at $z = 0.02$, which has however only a modest escape fraction. Another source showing both LyC emission and evidence of mergers is z19863 at $z = 3.088$ \citep[e.g.,][]{Riverathorsen17, Gupta2024}.
Finally, \cite{Maulick2024} reported LyC emission from a merging system at $z = 1.097$. Indeed, mergers can lead to inhomogeneous gas distributions, where tidal interactions strip gas and stars, resulting in complex morphologies \citep{Toomre1972, Cox2008, Pearson2019, Spilker2022}. These irregularities can result in anisotropic gas distributions, with pockets of optically thin neutral hydrogen allowing LyC photons to escape. In Haro 11, deep 21 cm observations have revealed reduced gas density pockets, likely due to recent merger activity, facilitating LyC photon escape \citep{LeReste2024}. 
Further supporting the link between mergers and LyC escape, \cite{Yuan2024} 
are the spatial offsets between the ionizing and non ionizing radiation observed in several sources such as the famous Sunburst arc \citep[$z = 2.4$, ][]{Riverathorsen17, Rivera-Thorsen2019, Mainali2022} where LyC emission originates from a leaking star-forming knot distinct from the rest of its host galaxy. Other examples are reported by \cite{Yuan2024} and \cite{Ji2020}.


The diversity in LCE morphologies highlights the importance of galaxy structure and gas distribution in determining LyC escape. In particular, the anisotropic coverage of neutral gas, rather than relying solely on simplified one-dimensional indicators like the O32 ratio -- which can also be influenced by variations in metallicity and ionization parameters \citep[e.g.,][]{Bassett2019, Katz2020} -- may play a crucial role in regulating the escape of LyC photons. Studies have shown that LCEs with disturbed morphologies often exhibit weaker correlations between the O32 ratio and $f_{\rm esc}$, suggesting that gas geometry and distribution are critical factors \citep{Bassett2019}.

With the advent of the \textit{James Webb Space Telescope} \citep[JWST, ][]{Gardner2023}, we now have the capability to probe the indirect indicators at unprecedented depths, enabling the measurement of $f_{\rm esc}$ for a large and diverse sample of galaxies during the EoR. Thanks to JWST's Near Infrared Camera (NIRCam), we are also refining our understanding of the morphology of these early sources \citep[e.g.,][]{Treu2023, Dalmasso2024}. Current findings indicate that signatures of mergers and interactions are detected in approximately 30\% of the galaxies studied at $z \geq 5.5$. Recent work by \cite{Calabro2024} showed that, in the GLASS and CEERS spectroscopic surveys, $\Sigma_{\text{SFR}}$ at $z \geq 7$ correlates with the O32 ratio and therefore with indirect estimates of the LyC escape fraction. 
Recent findings by \cite{Simmonds2024} 
demonstrate that galaxies during the EoR that are characterized by low stellar masses and bursty SFHs exhibit the highest values of $\xi_{\rm ion}$, driven by the dominance of younger stellar populations. This bursty star formation scenario is consistent with recent results from stacking analysis of the LzLCS sample \citep{Flury2024}, which suggest that such SFHs provide the optimal feedback and geometric conditions necessary for efficient LyC escape.

This work aims to explore in a systematic way the impact of galaxy morphology, star formation activity, and gas distribution on the physical processes driving LyC production and escape in galaxies during the EoR. 

The paper is structured as follows: in Sec.~\ref{sec:data}, we describe the sample selection. In Sec.~\ref{sec:method}, we present the methodology for line measurements, the SED fitting process, the general properties of the sources and the mergers identification criteria, while in Sec. \ref{sec:cox} we introduce a revised survival analysis to predict the $f_{\rm esc}$ for our sample. In Sec.~\ref{sec:result}, we report the predicted $f_{\rm esc}$, analyze the correlation of the presence of merger signatures with $f_{\rm esc}$ and $\xi_{\rm ion}$, and interpret our results. In Sec.~\ref{sec:conclusions}, we summarize our key findings. Throughout this work, we assume a flat $\Lambda$ cold dark matter cosmology with $H_0$ = 67.7 km s$^{-1}$ Mpc$^{-1}$ and $\Omega_m$ = 0.307 \citep{Planck2020} and the \cite{Chabrier_03} initial mass function. All magnitudes are expressed in the AB system \citep{Oke1983}.

\section{Data}\label{sec:data}

\begin{table*}[ht!]
\centering
\caption{Summary of JWST datasets and the count of confirmed sources with redshifts $5 \leq z \leq 7$ included in this study.}\label{tab:datasets_summary}
\begin{tabular}{lllll}
\hline\hline
Prog. name & PI & Target & Prog ID & $N_{\text{sources}}$ \\
\hline
EIGER & Lilly & SDSS J010013.02+280225.8 & GTO 1243 & 133 \\
CEERS & Finkelstein & CANDELS-EGS & ERS 1345 & 28 \\
DD-2750 & Arrabal Haro & CANDELS-EGS & DDT 2750 & 15 \\
JADES & Eisenstein & GOODS-S & GTO 1180 & 90 \\
JADES & Eisenstein & GOODS-N & GTO 1181 & 108 \\
JADES & Eisenstein & GOODS-S & GO 3215 & 24 \\
JADES & Luetzgendorf & GOODS-S & GTO 1210 & 30 \\
JADES & Luetzgendorf & GOODS-S & GTO 1286 & 8 \\
\hline
\end{tabular}
\end{table*}

\subsection{EIGER data}

We selected 133 sources from the Emission-line Galaxies and Intergalactic Gas in the Epoch of Reionization (EIGER) survey \citep[GTO 1243, PI: Lilly,][]{Kashino2023}. This study leverages the first deep 3.5 $\mu$m JWST/NIRCam Wide Field Slitless Spectroscopy (WFSS) observations from the EIGER program, focusing on spectroscopically confirmed [\textrm{O}\textsc{iii}] emitters at redshifts $z = 5.33 - 6.93$. These sources are located in the field of the bright quasar SDSS J010013.02+280225.8, though the majority are not directly associated with the quasar. A comprehensive list of identified sources and detection methods is available in \cite{Kashino2023, Matthee2023}. The dataset is further complemented by EIGER NIRCam imaging in three broadband filters (F115W, F200W, and F356W). For details on the imaging data reduction and analysis, we refer to \cite{Kashino2023} and \cite{Matthee2023}.

Our focus is on sources in the redshift range $5 \leq z \leq 7$, which were identified using NIRCam WFSS in the F356W filter, where the [\textrm{O}\textsc{iii}] and H$\beta$ emission lines are detectable. For point sources, the spectral resolution is approximately $R \sim 1500$ at 3.5 $\mu$m, with a dispersion of about 1 nm per pixel.


Emission lines were detected with a minimum signal-to-noise ratio (S/N) of 3, with a 3$\sigma$ limiting sensitivity varying across the field and with wavelength (by a factor of approximately 2), reaching $0.6 \times 10^{-18}$ erg s$^{-1}$ cm$^{-2}$ at 3.8 $\mu$m. After visual inspection and parameter refinement, 133 resolved [\textrm{O}\textsc{iii}] emitting components within the redshift range $z = 5.33 - 6.93$ have been identified, each with at least two detected emission lines at S/N > 3. The typical S/N for the [\textrm{O}\textsc{iii}]$\lambda 5008$ line is 14 (ranging from 6 to 70), while H$\beta$ is detected with S/N > 3 in 68 objects, with 31 of those objects reaching S/N > 5. Only 3 of the 133 objects were primarily identified via H$\beta$, where [\textrm{O}\textsc{iii}]$\lambda$4960 had S/N < 3.

\begin{figure}[t!]
\centering
\includegraphics[width=\linewidth]{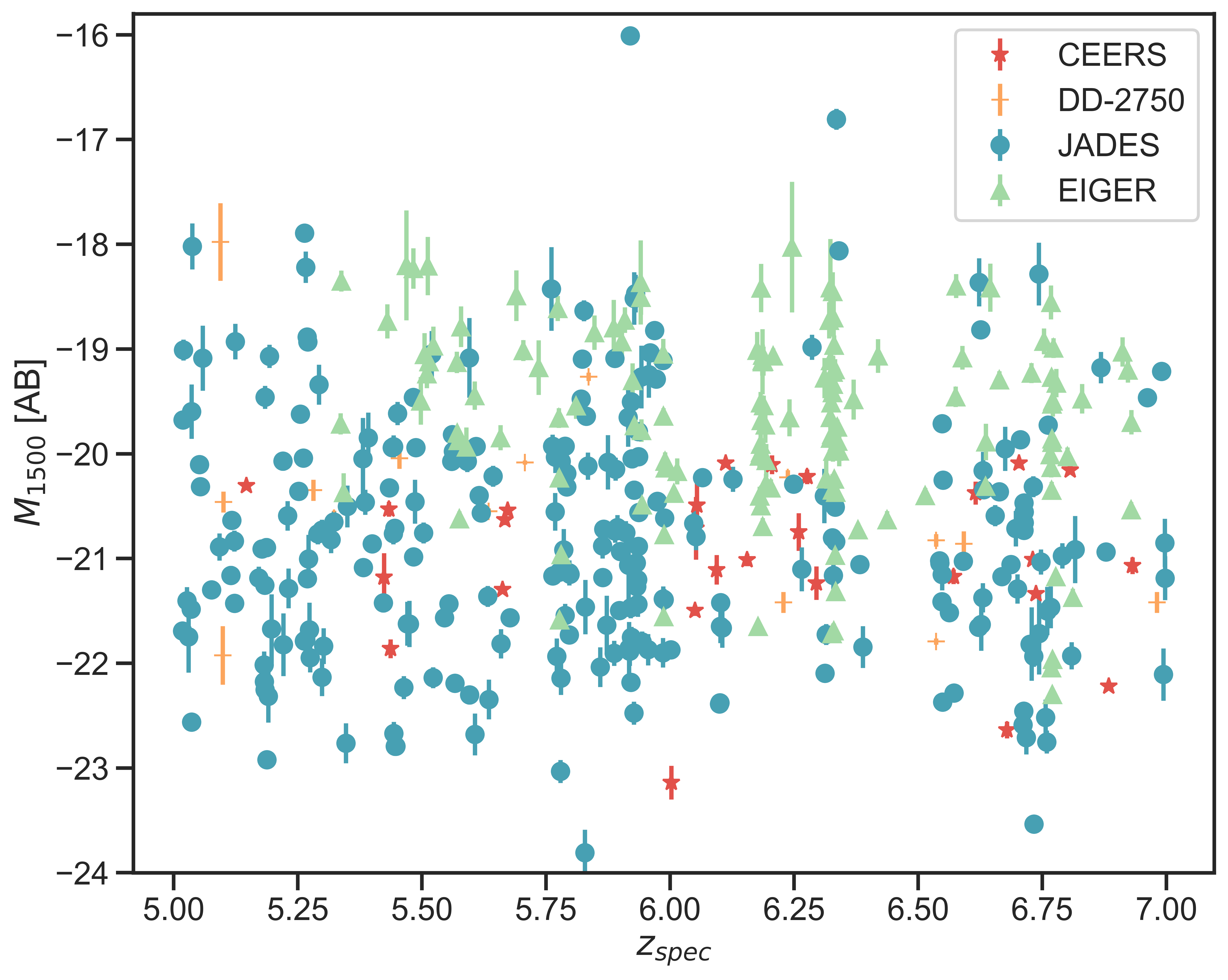}
\caption{$M_{1500}-z$ distribution for the 436 sources at $5 \leq z_{spec} \leq7$ analyzed in this work.}\label{fig:muv-z}
\end{figure}

\subsection{CEERS and DD-2750}
We selected sources from the Cosmic Evolution Early Release Science survey (CEERS; ERS 1345, PI: Finkelstein) in the Extended Groth Strip (EGS) field of CANDELS \citep{Grogin2011, Koekemoer2011}. In particular we 
selected all the sources with $5\leq z_{spec} \leq 7$ from \cite{Napolitano2024, Mascia2024_CEERS} that have a NIRSpec spectrum obtained either with the three medium-resolution ($R \approx 1000$) grating spectral configurations (G140M/F100LP, G235M/F170LP and G395M/F290LP), which, together, cover wavelengths between 0.7-5.1 $\mu$m, or with the PRISM/CLEAR configuration, which provides continuous wavelength coverage of 0.6-5.3 $\mu$m with spectral resolution ranging from $R \sim 30$ to 300. The total number of sources is 28.

All sources have available photometry with JWST/NIRCam filters (F115W, F150W, F200W, F277W, F356W, F444W, F410M), as well as HST/ACS filters (F606W, F814W) and HST/WFC3 filters (F105W, F125W, F140W, F160W) \citep{Bagley2023, Finkelstein2022, Finkelstein2022b}. To maintain consistency with the EIGER sample, we applied a cut in the F115W magnitude at 28.5 AB during the selection process. 

In addition, we also considered sources from a DDT program on the same field (DDT 2750, PI: Arrabal Haro). These sources were observed using the PRISM/CLEAR configuration and possess the same photometric information as the sources from the CEERS sample. 
We determined the redshifts of the galaxies by analyzing the 1D spectrum with the redshift-fitting module of \textsc{MSAEXP}\footnote{\url{https://github.com/gbrammer/msaexp}}. This module assesses the goodness of fit across a redshift range of 0 to 15, employing a series of \textsc{EAzY} \citep{Brammer2021} galaxy and line templates at the nominal NIRSpec/PRISM resolution. Each galaxy was fitted individually, and the resulting solutions were visually inspected. Similar to the CEERS sources, during the selection phase we implemented a cutoff for the F115W magnitude at 28.5 AB. In total we added 15 sources from this sample, in the redshift range of interest.

\begin{figure*}[ht!]
\centering
\includegraphics[width=\linewidth]{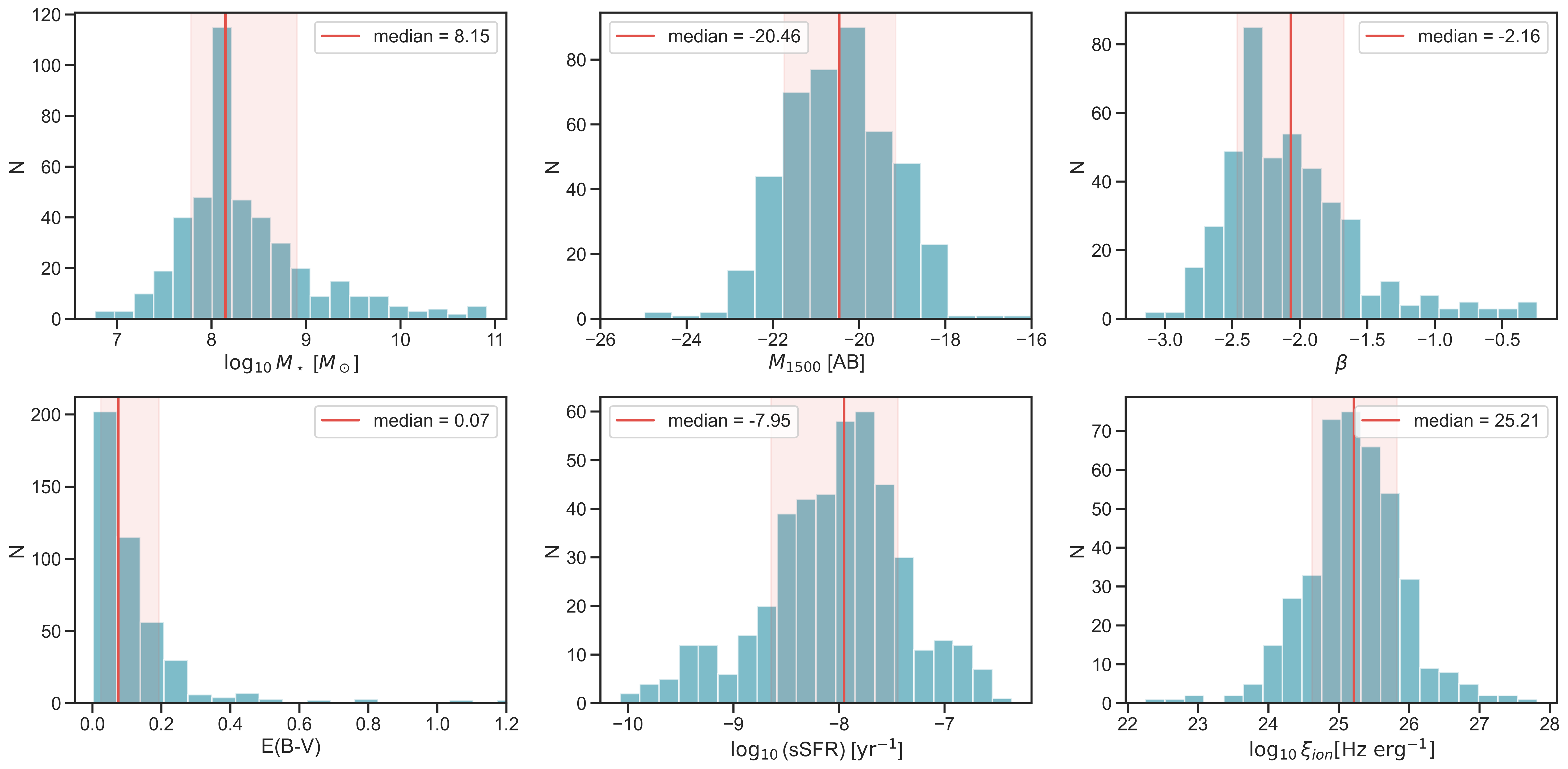}
\caption{Stellar mass $\log_{10}(M_\star)$, UV magnitude $M_{1500}$, $\beta$ slope, E(B-V), and $\log_{10}(\text{sSFR})$ for 436 sources in our sample, derived using \texttt{Prospector}. The ionizing photon production efficiency, $\log_{10} \xi_{\rm ion}$, has been calculated for all sources with detectable H$\alpha$ or H$\beta$ using the relations proposed by \cite{Leitherer1995, Schaerer2016}. The figure includes the median values along with their standard deviations (16th and 84th percentiles) for each distribution.}\label{fig:prop_total_sample}
\end{figure*}

\subsection{JADES}
We selected 
sources from the JWST Advanced Deep Extragalactic Survey \citep[JADES, GTO 1180, GTO 1210, GTO 1286 and GO 3215,][]{Eisenstein2023} in the Great Observatories Origins Deep Survey \citep[GOODS,][]{Giavalisco2004} South field and 
in the GOODS North field. All sources have available photometry with JWST/NIRCam filters (F115W, F150W, F200W, F277W, F356W, F444W, F410M), as well as HST/ACS filters (F606W, F814W) and HST/WFC3 filters (F105W, F125W, F140W, F160W) \citep{Merlin2024}.

We restricted the sample to sources with a spectroscopic redshift $5\leq z \leq 7$ from the Data Release 3 \citep{dEugenio2024}. The sources have a NIRSpec spectrum obtained either with the three medium-resolution grating spectral configurations (G140M/F100LP, G235M/F170LP and G395M/F290LP), or with the PRISM/CLEAR configuration. We included all sources within the specified redshift range flagged as ``A'', ``B'', or ``C'', indicating emission lines detected in the grism or prism configuration, respectively, or secure redshifts visually identified from spectral breaks and/or low signal-to-noise ratio (S/N) emission lines.
The final number of sources selected is 152 in the GOODS-S field and 108 in the GOODS-N field.

The list of the data sets along with their targeted fields is presented in Table~\ref{tab:datasets_summary}. 
When combined with the other samples, our final dataset comprises a total of 436 sources observed during the EoR. Fig.~\ref{fig:muv-z} illustrates the distribution of our total sample across the $z-M_{1500}$ plane. All $M_{1500}$ values were computed using \texttt{Prospector} (see Sec.~\ref{sec:method}).

\section{Methods}\label{sec:method}

\subsection{SED fitting}

The physical properties of the EIGER sources, 
presented in \cite{Matthee2023}, have been estimated using the \texttt{Prospector} code \citep{Johnson2021}. To ensure consistency in the derivation of physical properties across our entire sample, we utilized the same code, incorporating a nebular treatment based on Cloudy version 13.03 \citep{Ferland1998, Byler2017}, to fit the SEDs of sources from the other public programs. This approach allowed us to fit the available multiband JWST and HST photometry uniformly across the dataset.

Consistent with the methodology outlined by \cite{Matthee2023}, the free parameters in our modeling included the total formed stellar mass, stellar metallicity, star formation history, dust attenuation, gas-phase metallicity, and ionization parameter. The redshift was fixed based on the spectroscopic measurement, and we assumed a \cite{Chabrier_03} initial mass function along with MIST isochrones \citep{Dotter2016, Choi2016}. The star formation history followed a delayed-$\tau$ model, $\psi(t) = \psi_0 t e^{-t/\tau_T}$. Both stellar and nebular emissions were attenuated by dust using a simple screen model with a \cite{Calzetti2000} law.

We employed uniform and broad priors in our model: the stellar mass was allowed to vary between $10^{6}$ and $10^{11} \, M_{\odot}$, stellar metallicity between $[\text{Z}/\text{H}] = -2.0$ and $+0.2$, dust optical depth $\tau$ between 0 and 2, stellar age between 1 Myr and the age of the Universe at the galaxy's redshift, and the star formation timescale $\tau_T$ between 100 Myr and 20 Gyr. Gas-phase metallicity was modeled in the range $12 + \log(\text{O}/\text{H}) = 6.7$ to 9.2, while the ionization parameter $U$ spanned from $-3$ to $+1$ on a logarithmic scale, allowing for higher values to provide greater flexibility in the fits. 

The distribution of the main properties of the sources in this sample is presented in Fig.~\ref{fig:prop_total_sample}. The resulting SED fits suggest that in our galaxies dust attenuation is minimal, with a median $E(B-V) = 0.07$. The galaxies' observed UV magnitudes range from $M_{\rm UV} = -16.1$ to $-24.6$ (with a typical $M_{\rm UV} = -20.46$), and their stellar masses span three orders of magnitude, from $\log_{10}(M/M_{\odot}) = 6.76$ to 10.9, with a median mass of $10^{8.15} M_{\odot}$. 

\subsection{\texorpdfstring{Line measurements and $\xi_{\rm ion}$ estimation}{Line measurements and xi ion estimation}}

The catalog of emission lines detected in the EIGER sample is available \citep{Kashino2023}, providing fluxes and equivalent widths (EWs) for the H$\beta$ and [\textrm{O}\textsc{iii}] doublet lines.

 For the remaining samples, we measured emission lines using \textsc{LiMe}\footnote{\url{https://lime-stable.readthedocs.io/en/latest/}} \citep{Fernandez2024}, a versatile tool for analyzing medium-resolution and prism/clear 1D spectra. The tool requires the source’s 1D spectrum, spectroscopic redshift, and a predefined list of lines to fit. Our analysis focused on key emission lines, including [\textrm{O}\textsc{ii}], H$\beta$, [\textrm{O}\textsc{iii}], and H$\alpha$. \textsc{LiMe} estimates the continuum and computes direct integration and Gaussian fluxes for lines with a signal-to-noise ratio (S/N) above 3, as well as their EWs.

Before any quantitative analysis, we corrected the line fluxes for dust reddening. Since H$\alpha$ is unobserved for most sources, we used stellar E(B-V) values derived from SED fitting to calculate the dust attenuation for each emission line, applying the \cite{Calzetti2000} extinction law. The total attenuation in magnitudes was given by $A_V = 0.44 \times \text{E(B-V)} \times \kappa$, where $\kappa$ is the wavelength-dependent attenuation coefficient for each line. The corrected fluxes were obtained by multiplying the observed fluxes by $10^{A_V}$, with $\kappa_{\rm H\beta} = 4.60$, $\kappa_{\rm [O\textsc{iii}]} = 4.46$, and $\kappa_{\rm [O\textsc{ii}]} = 5.86$ \citep{calzetti1997}. We then computed O32 line ratios and rest-frame EWs for the [\textrm{O}\textsc{iii}] doublet and/or H$\beta$.

Following \cite{Leitherer1995} and \cite{Schaerer2016}, we inferred $\log\xi_{\rm ion}$ from the dust-corrected H$\alpha$ or $H\beta$ luminosity, as these lines are clearly detected in 413 out of 436 sources (95\% of the sample). The distribution of $\log\xi_{\rm ion}$ is shown in Fig.~\ref{fig:prop_total_sample}, with a median $\log\xi_{\rm ion}$ of 25.21.

\begin{figure}[ht!]
\centering
\includegraphics[width=\linewidth]{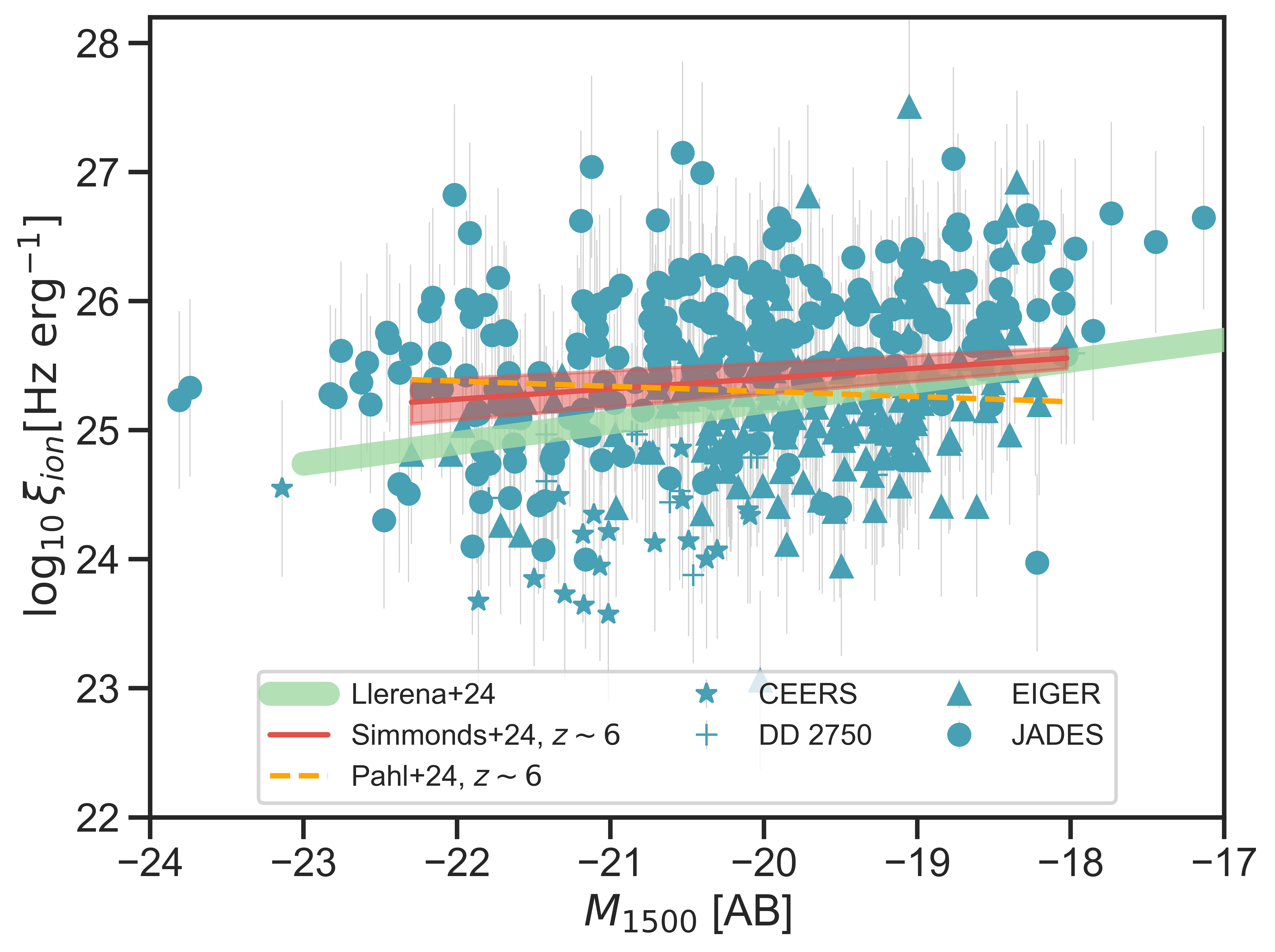}
\caption{$\log\xi_{\rm ion}$ versus $M_{\rm UV}$. Each data point corresponds to an individual galaxy from our sample, with different symbols representing the various surveys. We include the curves from \cite{Simmonds2024, Pahl2024}, as well as the relation from \cite{Llerena2024} at $z \sim 6$. } \label{fig:xiion_muv}
\end{figure}

In Fig.~\ref{fig:xiion_muv} we present the $\xi_{\rm ion}$ versus $M_{\rm UV}$. We observe that $\xi_{\rm ion}$ increases with $M_{\rm UV}$, consistent with the predictions at $z \sim 6$ from \cite{Simmonds2024, Prieto-Lyon2023, Llerena2024}. In our sample, fainter galaxies tend to have higher $\xi_{\rm ion}$ values compared to brighter galaxies, although there is considerable scatter. This finding is in contrast to recent results by \cite{Pahl2024}, which suggest that fainter galaxies are less efficient at producing ionizing photons than their brighter counterparts.

\begin{figure}[ht!]
\centering
\includegraphics[width=\linewidth]{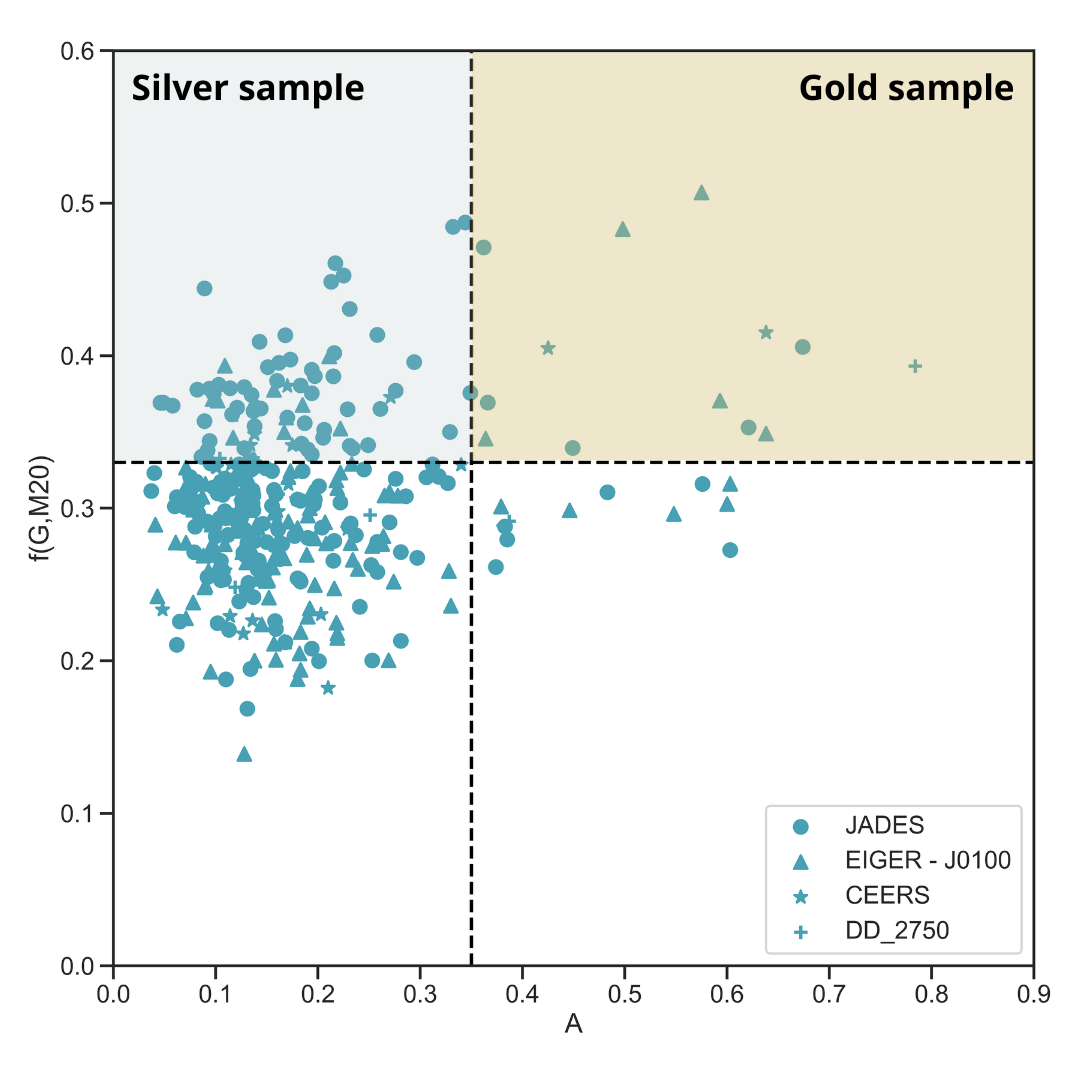}
\caption{Visual representation of our classification results in the $A$ vs $f(G, M_{20})$ diagram. Symbols are the same as in
Fig.~\ref{fig:xiion_muv}. The two dashed lines indicate the threshold values determined from Eqs. \eqref{eq:f} and \eqref{eq:A}.} \label{fig:merger_identification}
\end{figure}

\subsection{Mergers identification}

In this study, we aim to determine whether galaxy mergers create channels that facilitate the production and escape of LyC photons during reionization. To investigate this, we analyze NIRCam F115W images, available for all the surveys utilized in this work. These images trace the rest-frame UV at $\lambda_{\text{mean}} \sim 1916$ \AA\ for galaxies at $z = 5$ and $\lambda_{\text{mean}} \sim 1438$ \AA\ for galaxies at $z = 7$. Following \cite{Treu2023}, we create a segmentation map for each object to reduce noise and enhance low surface-brightness regions, particularly useful for galaxies with low signal-to-noise ratios \citep{Pawlik2016}. We apply a $6 \times 6$ uniform filter to the images and use the \texttt{photutils} package to generate the map, requiring at least 5 connected pixels with flux 2$\sigma$ above the background. The binary detection mask (MD) is defined as the segmentation area corresponding to the target, excluding neighbors. From MD, we derive the galaxy radius $R_{\rm max}$, the maximum pixel distance from the centroid \citep[more effective than the Petrosian radius for disturbed or low S/N morphologies,][]{Pawlik2016}. We set a lower limit of 0.025'' for spatial resolution, and structures below this scale are considered unresolved.

We also measure a set of morphological parameters, including the Gini coefficient \citep[$G$,][]{Abraham_2003}, which quantifies the distribution of light in a galaxy and indicates how concentrated or dispersed the light is; the second-order moment of brightness \citep[$M_{20}$,][]{Lotz_2004}, which measures the spatial concentration of the brightest 20\% of a galaxy's light; and asymmetry \citep[$A$,][]{Abraham_1996, Conselice_2000}, which assesses how the light distribution of a galaxy deviates from a perfectly symmetric shape. We successfully performed these measurements for 420 out of 436 sources (96\% of the sample). 

To identify mergers during the EoR, we apply two commonly used criteria from \cite{Conselice_2003} and \cite{Lotz_2008}: 

\begin{align} 
 f(G, M_{20}) = G + 0.14M_{20} &> 0.33 \label{eq:f}, \\
 A &\geq 0.35 \label{eq:A}.
\end{align} 

These criteria, initially developed for galaxies at $z \lesssim 1.2$ \citep{Conselice2009a}, have been shown to work well at higher redshifts \citep[e.g.,][]{Conselice2009, Treu2023, Vulcani2023, Dalmasso2024, Costantin2024}. 
It is worth noting that these criteria are effective for identifying galaxies in the early stages of a merger but might be less suited for detecting low surface-brightness features associated with post-merger stages \citep{Pawlik2016}. However, it is not clear if the creation of low density channels in the ISM is associated to specific phases of the mergers, and it might well be that they are not formed in the early stages. This will be further discussed in the results section.

In accordance with \cite{Dalmasso2024}, we define as 
Silver sample the galaxies that satisfy the Gini-$M_{20}$ criterion (Eq.\ref{eq:f}), and as Gold sample the subset of galaxies that also meet the asymmetry criterion (Eq.\ref{eq:A}). 


The Gold sample identifies mergers with more pronounced asymmetry, indicating more disruptive events, while the Silver sample captures mergers based primarily on light concentration. 
The results of this classification are illustrated in Fig.~\ref{fig:merger_identification}, with the two dashed lines indicating the threshold values calculated from Eqs. \eqref{eq:f} and \eqref{eq:A}. 
To ensure the accuracy of the classifications, we conducted a thorough visual inspection of all sources in both the Gold and Silver samples, finding no ambiguities.

The distribution of merger and non-merger signatures in our sample offers important insights into the nature of galaxies during the EoR. Quantitatively, only 13 out of 436 galaxies are part of the Gold sample (merger fraction $f_m = 0.03 \pm 0.01$), and 78 belong to the Silver sample (merger fraction $f_m = 0.17 \pm 0.01$). Interestingly, there are also 18 galaxies showing highly asymmetric shapes, but that are not selected as mergers. The vast majority -- 311 galaxies -- are compact systems. This distribution indicates that during the EoR, galaxies tend to be more compact rather than have major morphological disturbances. 

Our findings are consistent with previous studies. For instance, \cite{Dalmasso2024} reported a merger fraction for sources brighter than $M_{\rm UV} = -20.1$ for the Gold sample at $z \sim 6$ of $f_m = 0.06 \pm 0.03$ and for the Silver sample of $f_m = 0.22 \pm 0.04$, based on high-resolution NIRCam JWST data in the low-to-moderate magnification ($\mu$ < 2) regions of the Abell 2744 cluster field. 
Additionally, their analysis revealed no significant differences in specific star formation rates between merger and non-merger signatures, reinforcing the idea that interactions may not play a major role in regulating star formation during this period.
\section{\texorpdfstring{A revised model to infer $f_{\rm esc}$ using low-redshift analogs}{A revised model to infer fesc using low-redshift analogs}}\label{sec:cox}

\subsection{Identifying LzLCS+ analogs of EoR sources}
The LyC emitters from the LzLCS+ survey exhibit a wide range of physical properties, making them a diverse sample for studying LyC escape \citep{Flury2022b}. Their stellar masses vary from approximately $10^{7.2} M_\odot$ to $10^{9.3} M_\odot$, and their $M_{\rm UV}$ range from $-21.3$ to $-18.3$. The sample also shows considerable variability in dust content, reflected by $\beta$ slopes spanning from $-2.7$ to $-1.6$ and reddening values, E(B-V), between 0.013 and 0.206. 
Because LzLCS+ galaxies were selected to cover diverse properties and test $f_{\rm esc}$ diagnostics, they span a much wider (and in some cases different) range of properties than those observed during the EoR. Therefore, only a fraction of them can be considered true analogs of the sources currently detected by JWST in that era.

To account for these differences, we decided to refine our predictions for $f_{\rm esc}$ during reionization by comparing our sample of galaxies at $5 \leq z \leq 7$ with the LzLCS+ sample, with the aim of calibrating the low redshift models only on true analogs of EoR galaxies.

\begin{figure}[ht!]
\centering
\includegraphics[width=0.95\linewidth]{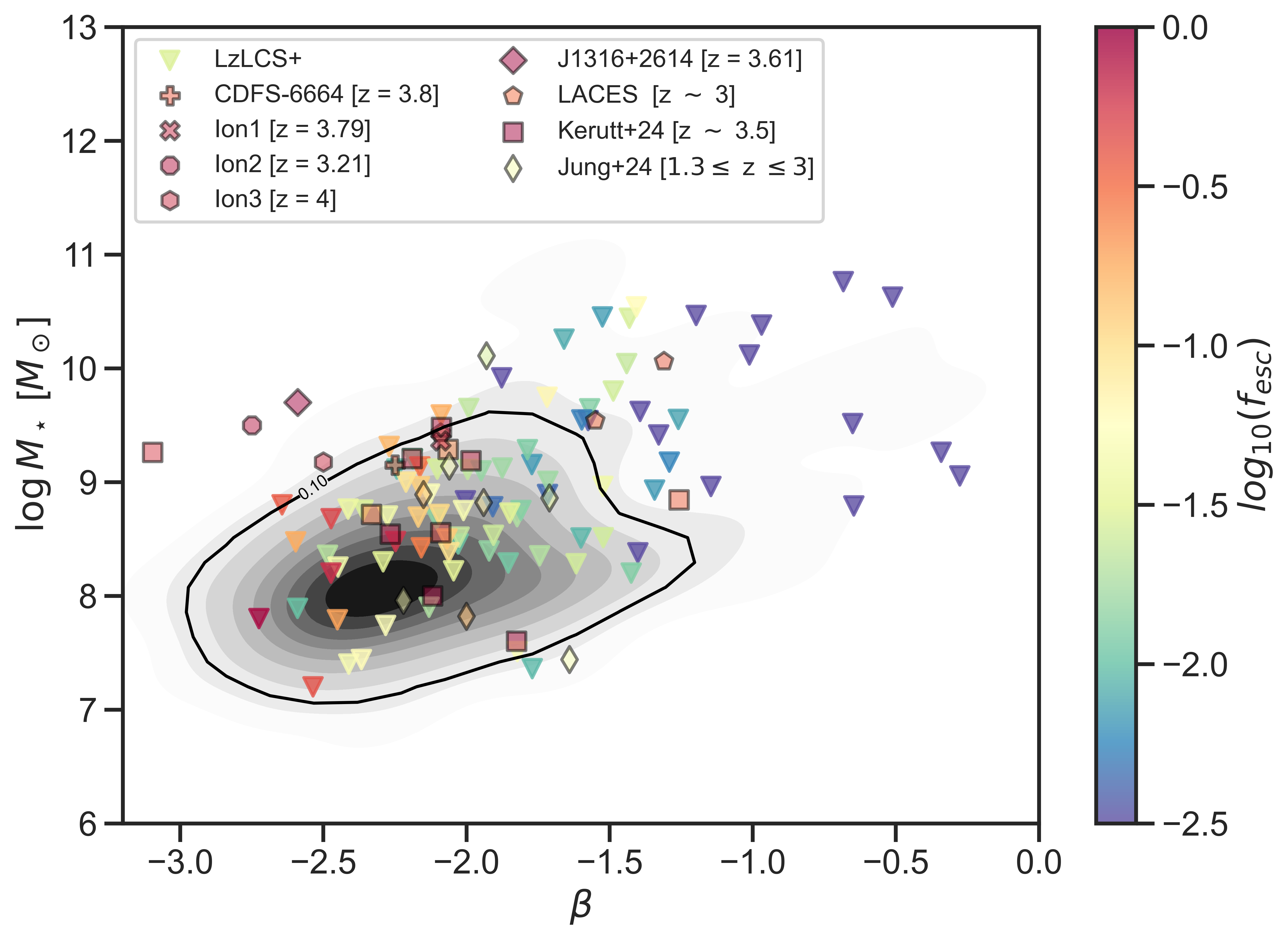}
\includegraphics[width=0.95\linewidth]{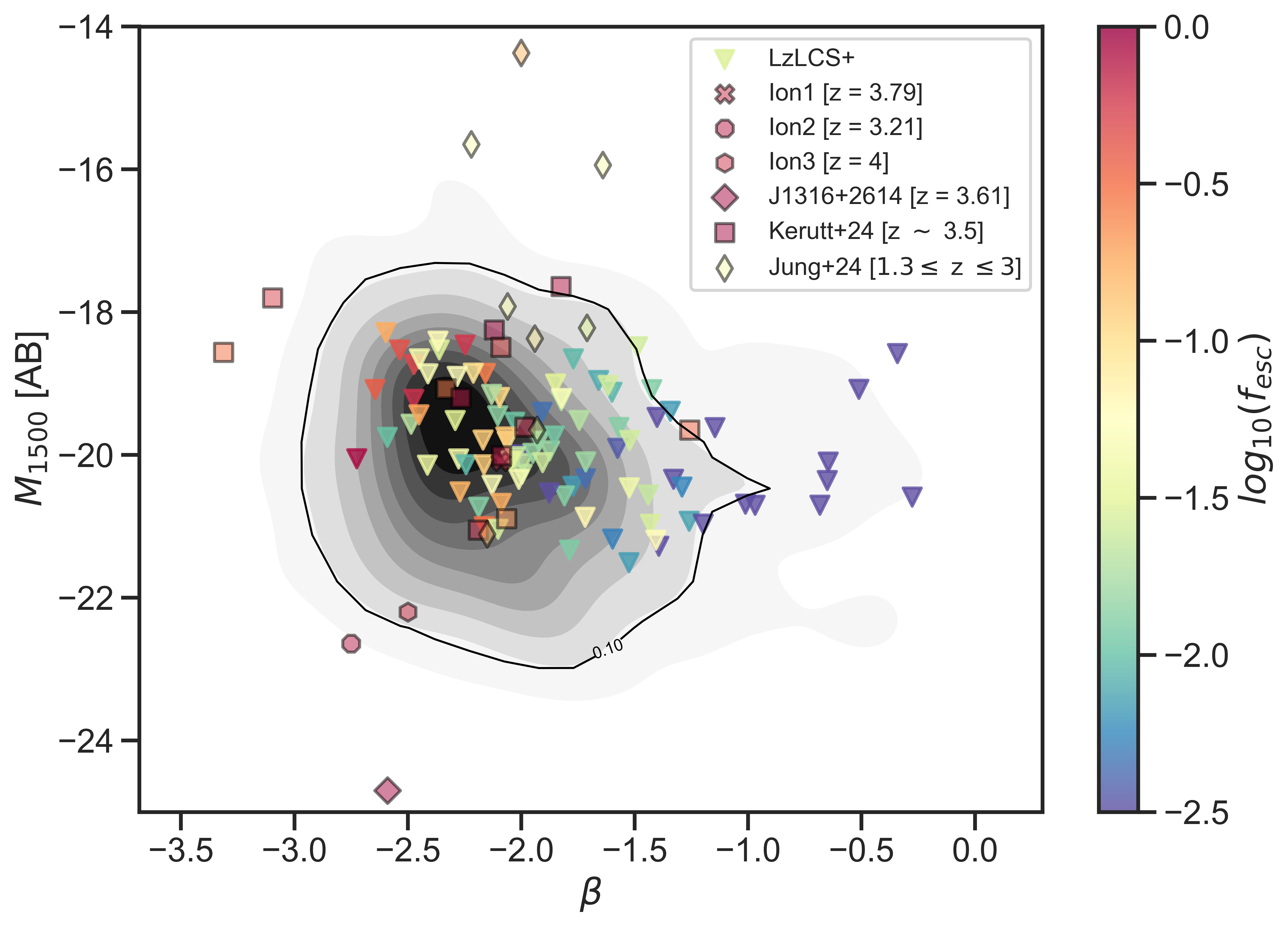}
\includegraphics[width=0.95\linewidth]{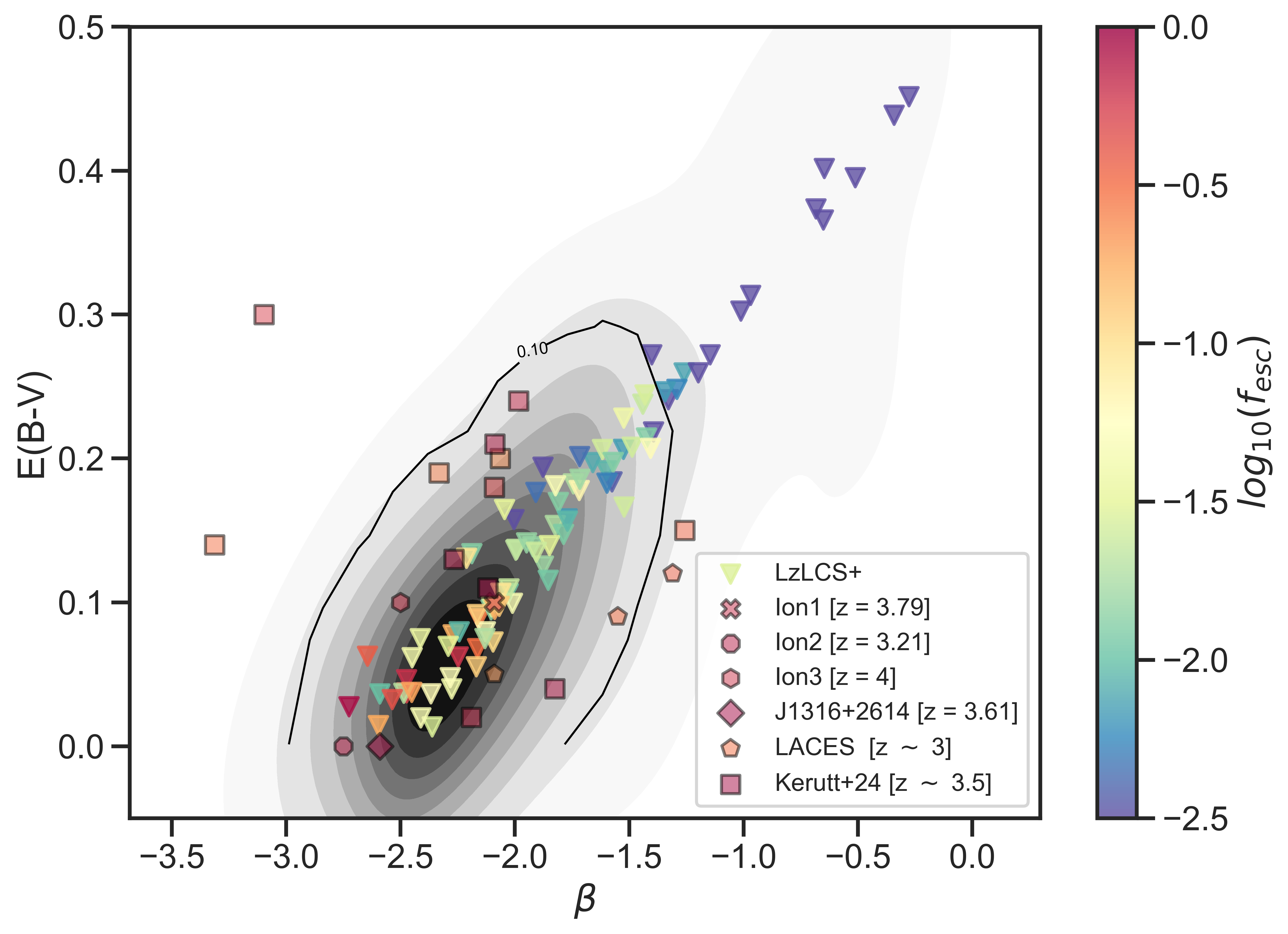}
\caption{Comparison of galaxy properties between $z \sim 0.3$ LyC emitters from the LzLCS+ \citep{Flury2022b} and high-redshift sources, shown as a density plot. Panels show $\log M_\star$ vs. $\beta$ slope (upper), $M_{\text{UV}}$ vs. $\beta$ (middle), and E(B-V) vs. $\beta$ (lower). Of the 88 LyC emitters, 51 galaxies are identified as analogs to high-redshift sources, falling within the 99\% confidence interval in all panels. We also include known $z \sim 3$ leakers from the literature: Ion1 \citep{Vanzella2012}, Ion2 \citep{Vanzella2015, Vanzella2016, deBarros2016}, Ion3 \citep{Vanzella2018}, J1316+2614 \citep{Marques-Chaves2022}, CDFS-6664 \citep{Yuan2021}, LACES \citep{Fletcher19}, and samples from \cite{Kerutt2024} and \cite{Jung2024}. All the known leakers are color-coded by their log$_{10}(f_{\rm esc})$.} \label{fig:analogs_prop}
\end{figure}

In Fig.~\ref{fig:analogs_prop}, we show the comparison between various properties of the galaxies, including $\beta$ slope vs. stellar mass, $\beta$ vs. $M_{\rm UV}$, and $\beta$ vs. dust reddening E(B-V). The grey-shaded area represents the density distribution of the sources in our EoR sample, with the contours in each panel indicating the 99\% interval of the distributions, while the individual points are the LzLCS+ sources Based on the four properties shown in the figure, we find that only 51 galaxies in the LzLCS+ sample match the properties of EoR sources, as they consistently fall within the 99\% distribution in each panel of the figure. From here on we refer to this subsample as ``analogs''. The mean properties indicate that the analogs are dust-poor (E(B-V) = 0.1), have a blue spectral slope ($\beta = -2.1$), a high O32 ($\sim 9$), a low stellar mass ($\log M_\star = 8.6 M_\odot$), a $M_{\rm UV}$ of approximately $-19.65$, and are compact in the UV ($r_e \sim 0.559 \ \text{kpc}$).
Additionally, in the plots we also show known LyC leakers at $z \sim 3$ from the literature \citep[e.g.,][]{Vanzella2012,Vanzella2015, Vanzella2016, Vanzella2018, deBarros2016, Fletcher19, Yuan2021, Marques-Chaves2022, Kerutt2024, Jung2024}, highlighting that even at this redshift, these galaxies exhibit a wide range of properties, not all of which make them analogs of EoR.

\begin{figure}[ht!]
\centering
\includegraphics[width=0.9\linewidth]{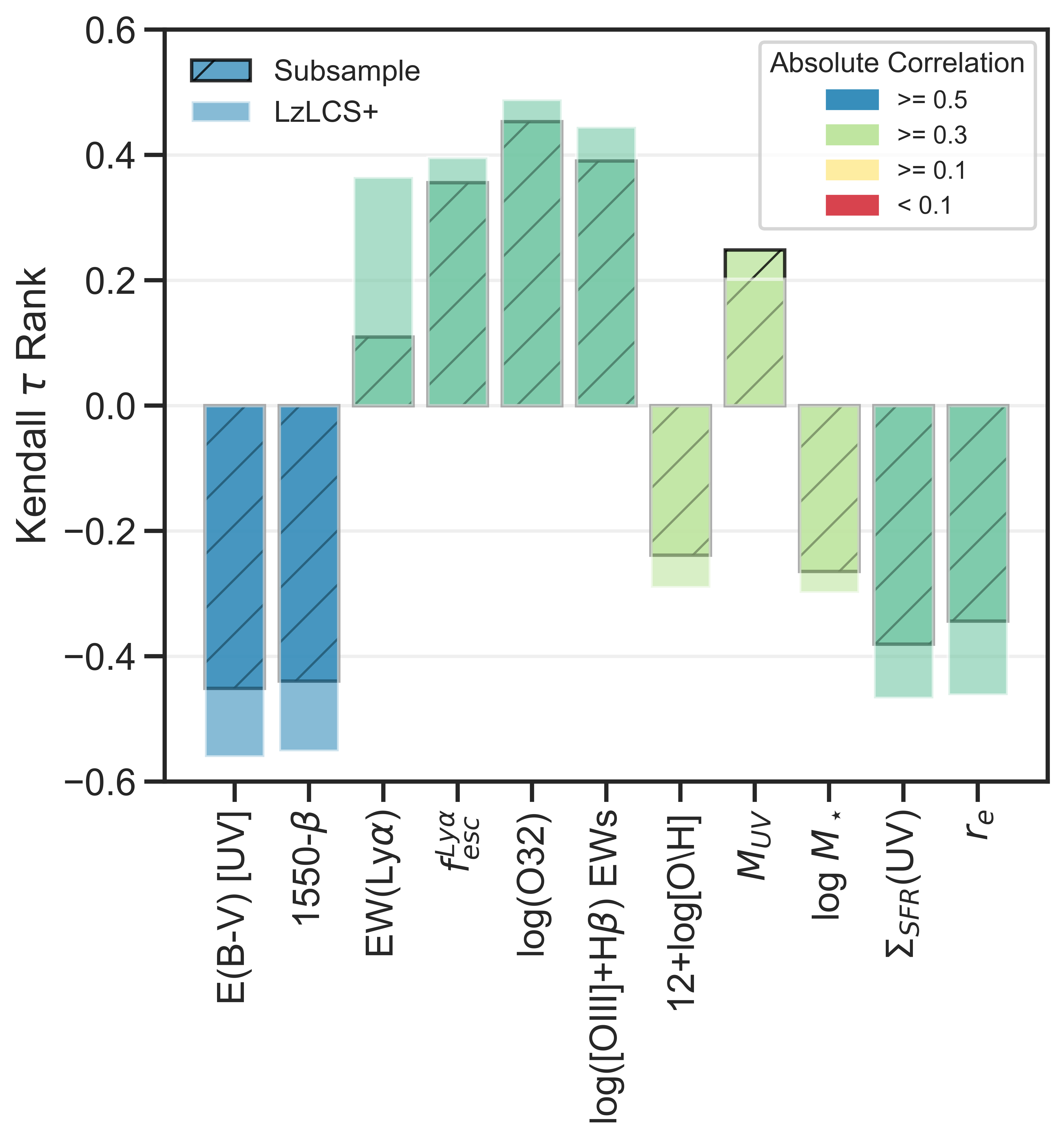}
\caption{Results of the Kendall $\tau$ rank correlation analysis between various predictors identified by \cite{Jaskot24b} and $f_{\rm esc}$ values for both the entire LzLCS+ sample (88 sources) and the subsample of 51 galaxies. The hatched bar represents the refined subsample, while the other corresponds to the full sample.} \label{fig:correlations}
\end{figure}

In Fig.~\ref{fig:correlations} we present the results of a Kendall $\tau$ rank correlation analysis conducted to evaluate the relationship between the various indirect indicators identified by \cite{Jaskot24b} and $f_{\rm esc}$ values from the LzLCS+ sample.
We followed the approach outlined by \cite{Flury2022}\footnote{\url{https://github.com/sflury/kendall}}, computing the Kendall $\tau$ rank correlation coefficient using the \cite{Akritas1996} method, which accounts for censored data. 
We perform this analysis both for the entire sample of 88 LzLCS+ sources and for the 51 analogs. The correlation 
between $f_{\rm esc}$ and most properties 
show only very minor variations, suggesting that their predictive power remains relatively stable across both samples. 
However, the correlation with $M_{\rm UV}$ becomes stronger, and the one involving the \lya\ EW shows the greatest variation, in the sense that using only the 51 analogs, \lya\ EW and $f_{\rm esc}$ do not show any correlation.

We first examined whether the reduction in sample size could explain the changes in Kendall $\tau$ rank correlation. To do this, we generated 200 random subsamples, each containing 51 sources (the same size as our subsample), and calculated the correlation coefficients for each. We found that, for most of the indirect indicators, the small decrease in correlation strength observed in the 51-source subsample can be attributed to the reduced sample size, suggesting that the weakening of these correlations is largely a statistical effect. However, the observed change in correlation strength for \lya\ EW is significantly larger than expected from sample size reduction alone. Furthermore, also the small increase in correlation with $M_{\rm UV}$ could not be explained by the reduced sample size. Therefore, the variations in $M_{\rm UV}$ and \lya\ EW correlations likely reflect real physical differences, rather than just the impact of a smaller sample size.

The absence of correlation with \lya\ EW is surprising, as this has always been claimed to be one of the most solid indirect indicators. However, this picture has been recently questioned: for example \cite{Citro2024}, explored the connection between LyC leakage and \lya\ emission in seven gravitationally lensed \lya\ emitters at $z \sim 2$. 
Their galaxies have similar properties to the LzLCS+ sample, 
and have high \lya\ $f_{\rm esc}$. Based on the relation calibrated on \lya\ properties using the LzLCS+ sample, at least half of these sources should be classified as LCEs but in reality, none of them is 
detected in LyC with $f_{\rm esc} \leq 0.065$. 
These findings suggest that the mechanisms governing LyC leakage may differ at high redshift, emphasizing the need for caution when applying low-redshift estimators to high-redshift sources. Indirect $f_{\rm esc}$ estimators like \lya\ EW may have complex dependencies beyond just LyC escape.
Given that galaxy properties evolve with redshift, it remains uncertain how these changes impact the diagnostic reliability for $f_{\rm esc}$.

\begin{figure*}[ht!]
\centering
\includegraphics[width=0.49\linewidth]{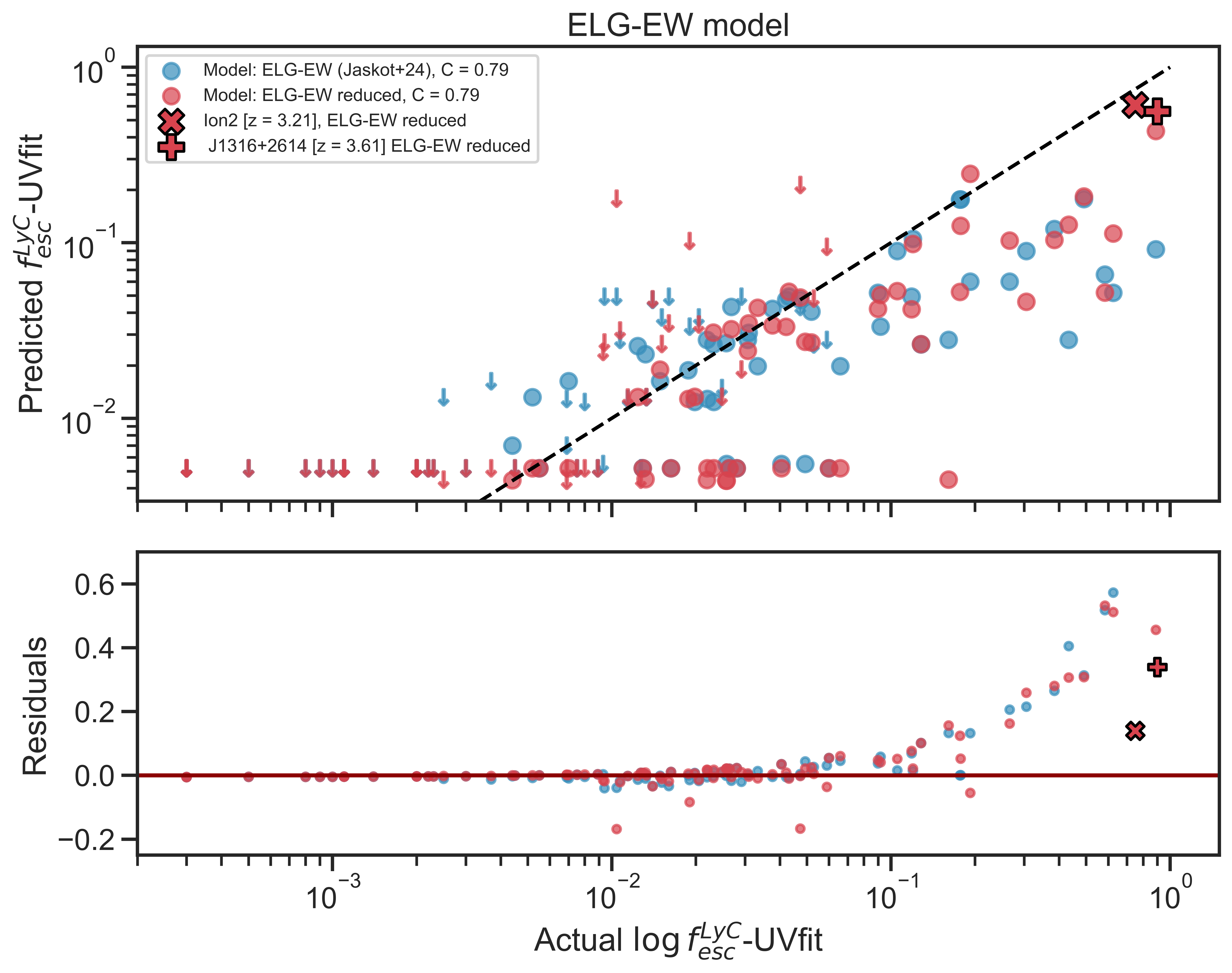}
\includegraphics[width=0.49\linewidth]{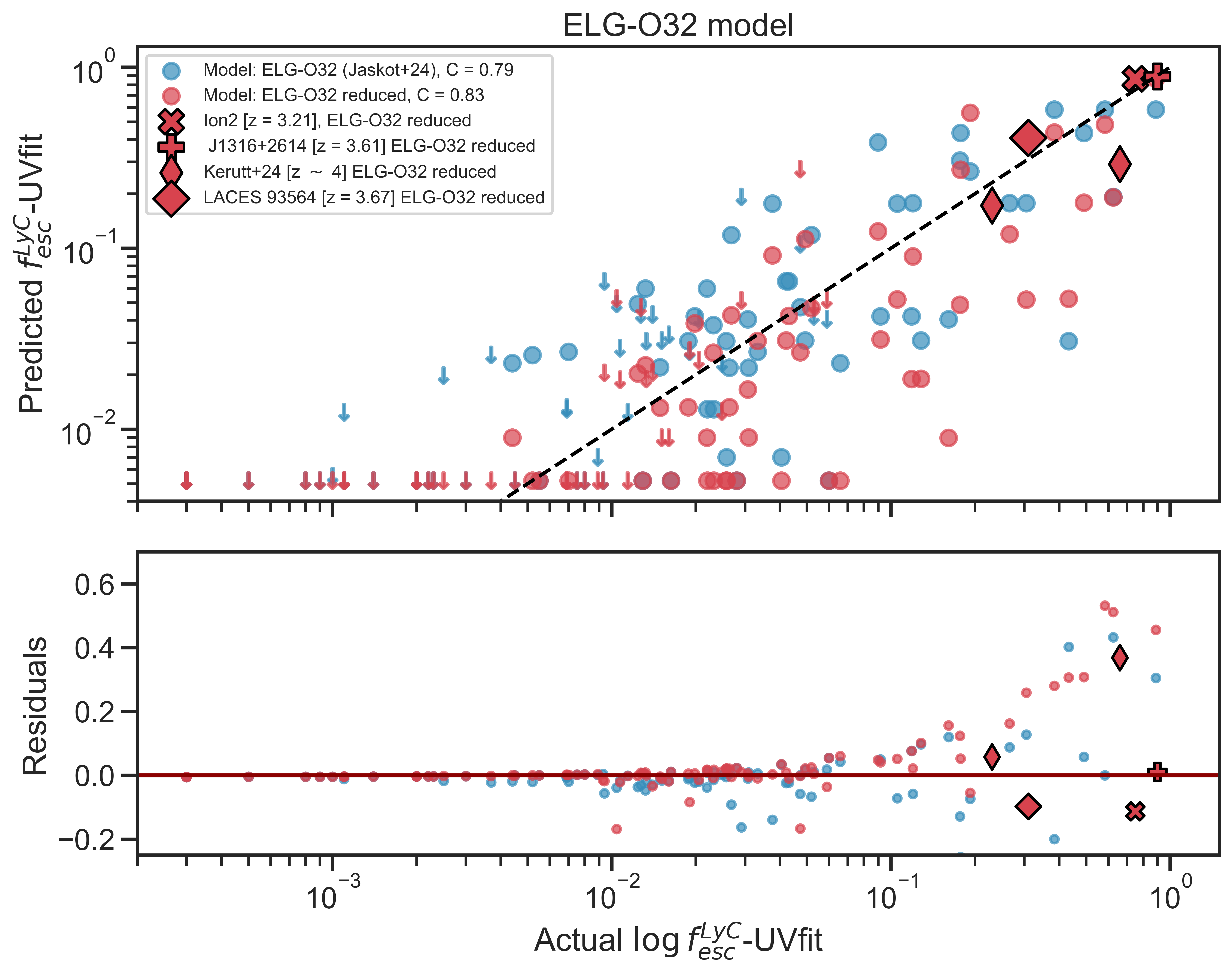}
\caption{Detected (or upper limit) $f_{\rm esc}$ values versus the predicted $f_{\rm esc}$ values using Cox models, calibrated using the full LzLCS+ sample and the subsample of galaxies resembling reionization-era sources. Left panel: Predictions using the ELG-EW model. Right panel: Predictions using the ELG-O32 model. The Cox models from \citep{Jaskot24b} are shown in blue, while the predictions calibrated on the subsample are in red. Predictions applied to known LyC leakers at $z \sim 3$ are also shown in red \citep{Vanzella2015, Vanzella2016, deBarros2016, Marques-Chaves2022, Fletcher19, Kerutt2024}. Upper limits are marked with downward-pointing arrows.} \label{fig:comparison_models}
\end{figure*}
\vspace{-0.2 cm}

\subsection{\texorpdfstring{A revised Cox model to predict $f_{\rm esc}$ during the EoR}{A revised Cox model to predict fesc during the EoR}}

Having identified the subset of LzLCS+ galaxies that exhibit properties similar to high-redshift sources, we recalibrate the Cox proportional hazards model -- a semi-parametric method that estimates the probability of an event (the detection of $f_{\text{esc}}$) as a function of multiple variables (the various galaxy properties) \citep{Jaskot24a, Jaskot24b} -- to more accurately predict $f_{\rm esc}$ during the EoR. This selection allows for a more reliable predictive framework, improving our ability to model the LyC escape properties of galaxies in the early Universe. In \nameref{appendixA}, we outline the methodology presented in \cite{Jaskot24a}.

The concordance index, defined as a measure to assess goodness-of-fit, ranges from 0 (perfect disagreement) to 0.5 (perfectly random) to 1.0 (perfect concordance) \citep[e.g.,][]{Davidson-Pilon2019}. 
From the models with a concordance coefficient $C > 0.7$ identified by \cite{Jaskot24b}, we select those applicable to reionization-era sources. Specifically, we focus on two models that: (a) can be applied to all sources in our sample, and (b) do not include a morphology-dependent term. The exclusion of a morphology term aligns with the goal of the present study, which is to investigate whether LyC escape is influenced by the morphology of galaxies and the presence of mergers with high $f_{\rm esc}$. Since all the models in \cite{Jaskot24b}, including those with a morphology term, are statistically equivalent, this selection does not introduce any significant bias. Thus, we choose the ELG-EW and ELG-O32 models, which rely on key galaxy properties such as E(B-V), $M_{\rm UV}$, stellar mass, and, respectively, $\log_{10}$([\textrm{O}\textsc{iii}]+H$\beta$) EWs and $\log_{10}$(O32). 

The ELG-O32 is our fiducial model, while the ELG-EW model is useful for sources where O32 cannot be computed due to the absence of one of the oxygen lines.
Following the methodology outlined in \nameref{appendixA}, we computed the baseline hazard function $H_{F_0}(f_{\text{abs}})$ and the survival function $S(f_{\text{abs}})$, which allowed us to estimate the median $f_{\text{abs}}$ (and consequently the median $f_{\rm esc}$), along with the 16th and 84th percentiles to establish confidence intervals for our predictions. 

In \nameref{appendixA}, we provide two tables (Tables~A.1 and~A.2), one for each model, which, following \cite{Jaskot24b}, provide the goodness-of-fit statistics for the subsample of LzLCS+ analogs, the fitted coefficients ($\beta_i$) for each included variable, and the reference values $\bar{x}_i$, which are the mean of the analogs $x_i$ values, where $x_i$ represents the input variable ($M_{\rm UV}$, E(B-V), $\log_{10}$($M_*$/$M_\odot$), and $\log_{10}$(EW([\textrm{O}\textsc{iii}]+H$\beta$)/\AA) or $\log_{10}$(O32)). Finally, the tables list the baseline cumulative hazard function, $H_{F_0}$, calculated by the \texttt{lifelines} Cox Model fitting routine \citep{Davidson-Pilon2019} for each of the observed $f_{\rm esc}$ values for LCEs in the LzLCS+ subsample.

Fig.~\ref{fig:comparison_models} illustrates the comparison between the predicted $f_{\rm esc}$ values using the full LzLCS+ sample and the analogs alone. Both predictions are consistent, with residuals showing similar trends across the two samples. 

\subsection{\texorpdfstring{Applying the model to known leakers at $z = 3$}{Applying the model to known leakers at z = 3}}

We further test our model to assess its capacity to predict $f_{\rm esc}$ for galaxies at intermediate redshift by applying it to known LyC leakers at $z \sim 3$. 
We compile samples of $z \sim 3$ leakers with reported detections of global absolute LyC $f_{\rm esc}$ values that also include all the relevant variables: E(B-V), stellar mass, $M_{\rm UV}$, and either O32 or [\textrm{O}\textsc{iii}]+H$\beta$ EWs. This results in two distinct sets at $z \sim 3$:
\begin{itemize}
 \item ELG-EW Sample: This sample includes Ion2 \citep[$z = 3.2$, $f_{\rm esc}$ = 0.5–0.9,][]{Vanzella2015, Vanzella2016, deBarros2016} and J1316+2614 \citep[$z = 3.6130$, $f_{\rm esc} \approx 0.9$,][]{Marques-Chaves2022, Marques-Chaves2024}.
 \item ELG-O32 Sample: This sample includes Ion2, J1316+2614, two sources from \cite{Kerutt2024} that have been observed in the JADES program ($z = 3.46$, $f_{\rm esc} = 0.23 \pm 0.05$ and $z = 4.43$, $f_{\rm esc} = 0.69 \pm 0.10$, respectively), and one source ($z = 3.67$, $f_{\rm esc} = 0.31 \pm 0.03$) from the Lyman Continuum Escape Survey \citep[LACES,][]{Fletcher19}.
\end{itemize}
Our predictions, shown in Fig.~\ref{fig:comparison_models}, 
revealed a strong positive correlation with measured values at this intermediate redshift, with Pearson's \textit{r }= 0.812, indicating that our model effectively reflects the observed data. Although the \textit{p}-value of 0.095 suggests marginal significance, this reinforces the validity of our predictions. 
A primary limitation of our assessment is that we are testing our model exclusively on strong leakers at $z \sim 3$ (with 
$f_{\rm esc} \geq 0.2$), as these are the only detections available with the complete dataset of indirect indicators needed. Consequently, expanding the number of confirmed LyC leakers (or upper limits) at intermediate redshifts is essential for further refinement of our model. 

\section{Results}\label{sec:result}

\subsection{\texorpdfstring{The predicted $f_{\rm esc}$ values}{The predicted fesc LyC values}}

\begin{figure}[ht!]
\centering
\includegraphics[width=\linewidth]{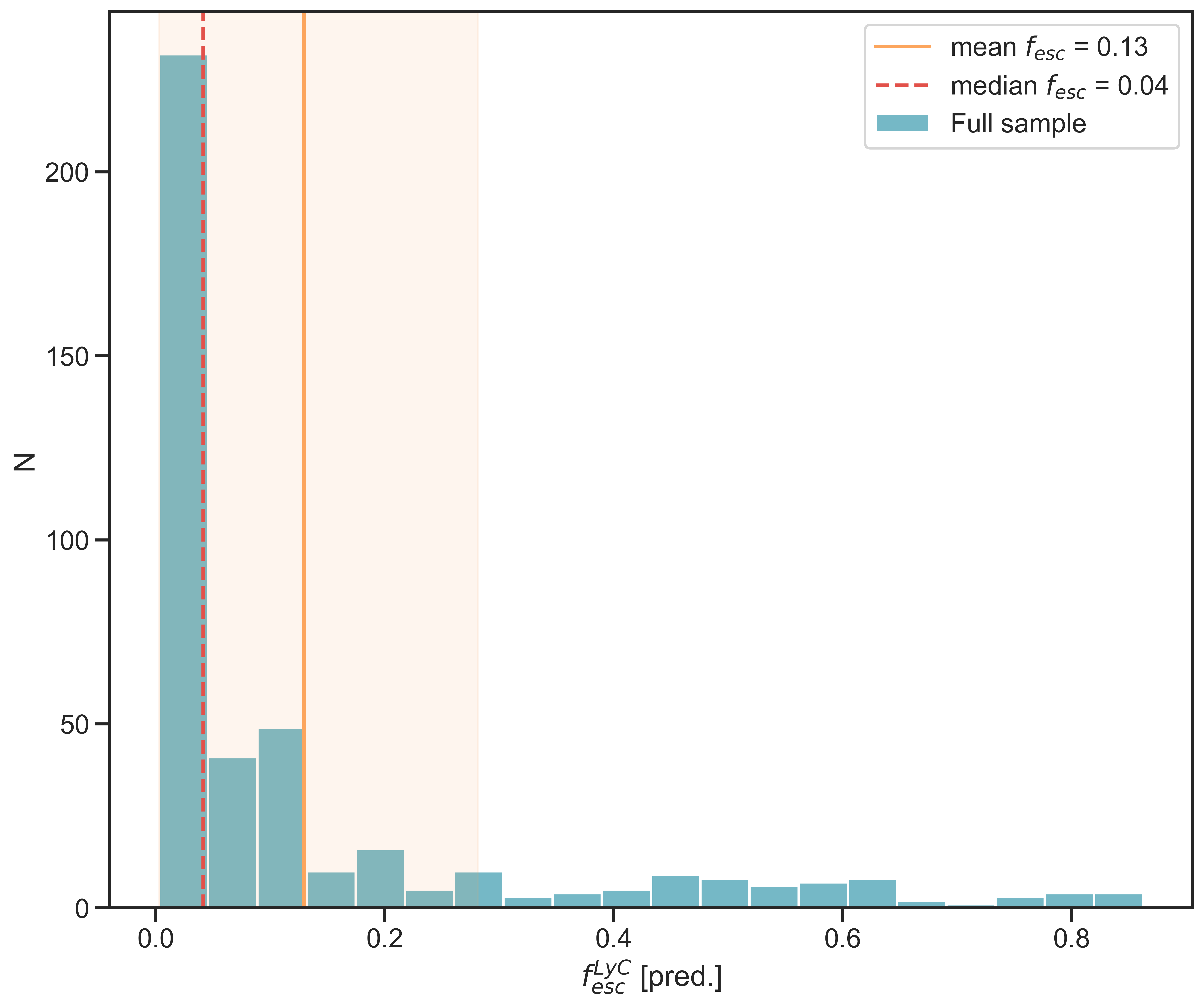}
\caption{Predicted $f_{\rm esc}$ distribution for the analyzed sources at $5 \leq z \leq7$. The mean $f_{\rm esc}$ of the sample is shown in red and it is equal to $0.13^{0.25}_{0.002}$, where the errors show the 16-84th percentiles. The median $f_{\rm esc} = 0.04$ is presented in orange.}\label{fig:fesc dist}
\end{figure}

Using the revised ELG-O32 or ELG-EW models, we predicted the $f_{\rm esc}$ values for most sources in our sample, successfully predicting all but 9 out of a total of 436 sources. The 9 sources for which predictions were not made lack both the O32 ratio and the [O\textsc{iii}] + H$\beta$ EW.
To obtain the most accurate estimates, we applied the revised ELG-O32 model (which has a concordance value of C = 0.83) whenever possible. For sources without detectable [O\textsc{ii}] or [O\textsc{iii}] lines, such as those in the EIGER sample, we applied the ELG-EW model (C = 0.79).
In particular, from the total sample of 427 sources with predicted $f_{\rm esc}$, we used the ELG-O32 model for 206 sources, and the ELG-EW models for the remaining 221 sources. To ensure consistency between the two methods, we tested the correlation between the predicted $f_{\rm esc}$ values from both models using the Spearman rank coefficient. We find a strong agreement ($r = 0.86, \text{p-value} = 0.003$). The distribution of predicted $f_{\rm esc}$ values is shown in Fig.~\ref{fig:fesc dist}. Most of our galaxies exhibit modest inferred $f_{\rm esc}$ values, generally around 0.10 or below. The average $f_{\rm esc}$ for the sample, including the standard deviation as the 16th and 84th percentile values of the distribution, is $0.13^{0.25}_{0.002}$. This mean is skewed by the relatively high $f_{\rm esc}$ values ($>0.4$) inferred for a modest fraction (53/427) of our sources. As a result, the median provides a more representative statistic, with a value of $0.04$. 

The results align well with previous findings \citep{Mascia2023_Glass, Mascia2024_CEERS}, both in terms of the shape of the distribution, mean, and median values. The models used in those studies were derived through linear regression, incorporating key predictors such as the O32 ratio or the H$\beta$ EW, the $\beta$ slope, and the UV half-light radius $r_e$. Our findings are also broadly consistent with \citet{Lin2023}, who applied a logistic regression model to estimate the probability of a galaxy with $M_{\rm UV} < -18$ being a LCE, using $M_{\rm UV}$ and $\beta$, and O32. They found that at $z = 8$, the average $f_{\rm esc}$ varies with $M_{\rm UV}$, peaking at intermediate luminosities ($-19 < M_{\rm UV} < -16$) and reaching about 0.04 for brighter galaxies ($M_{\rm UV} < -19$). 
Similarly, \cite{Saxena2024} observed a comparable trend in a smaller sample of Ly$\alpha$ emitters, where a few sources exhibit a high $f_{\rm esc}$ while the majority have $f_{\rm esc} \leq 0.10$. Their model, based on the $\text{\textsc{Sphinx}}^{20}$ simulation \citep{Choustikov2024}, used six parameters ($\beta$, E(B-V), H$\beta$ luminosity, M$_{\rm UV}$, R23, and O32) to predict LyC escape. It is important to note that all these methods are each subject to own caveats -- whether in the treatment of upper limits, assumptions in simulations, or the handling of non-detections as non-LCEs.
We note that the \cite{Choustikov2024} model, derived from simulated galaxies, substantially disagrees with the LzLCS+ sample, as shown by \cite{Jaskot24b}, and also diverges from Cox's predictions for $z \geq 6$ galaxies. Therefore, while these methods may yield similar population averages, they differ in their predictions for individual galaxies. Despite these differences, the methods demonstrate the strength of multivariate predictions and the value of combining different diagnostics to achieve a more robust understanding of LyC escape.

\subsection{\texorpdfstring{Merger fraction versus $f_{\rm esc}$}{Merger fraction versus fesc LyC}}

\begin{figure}[ht!]
\centering
\includegraphics[width=\linewidth]{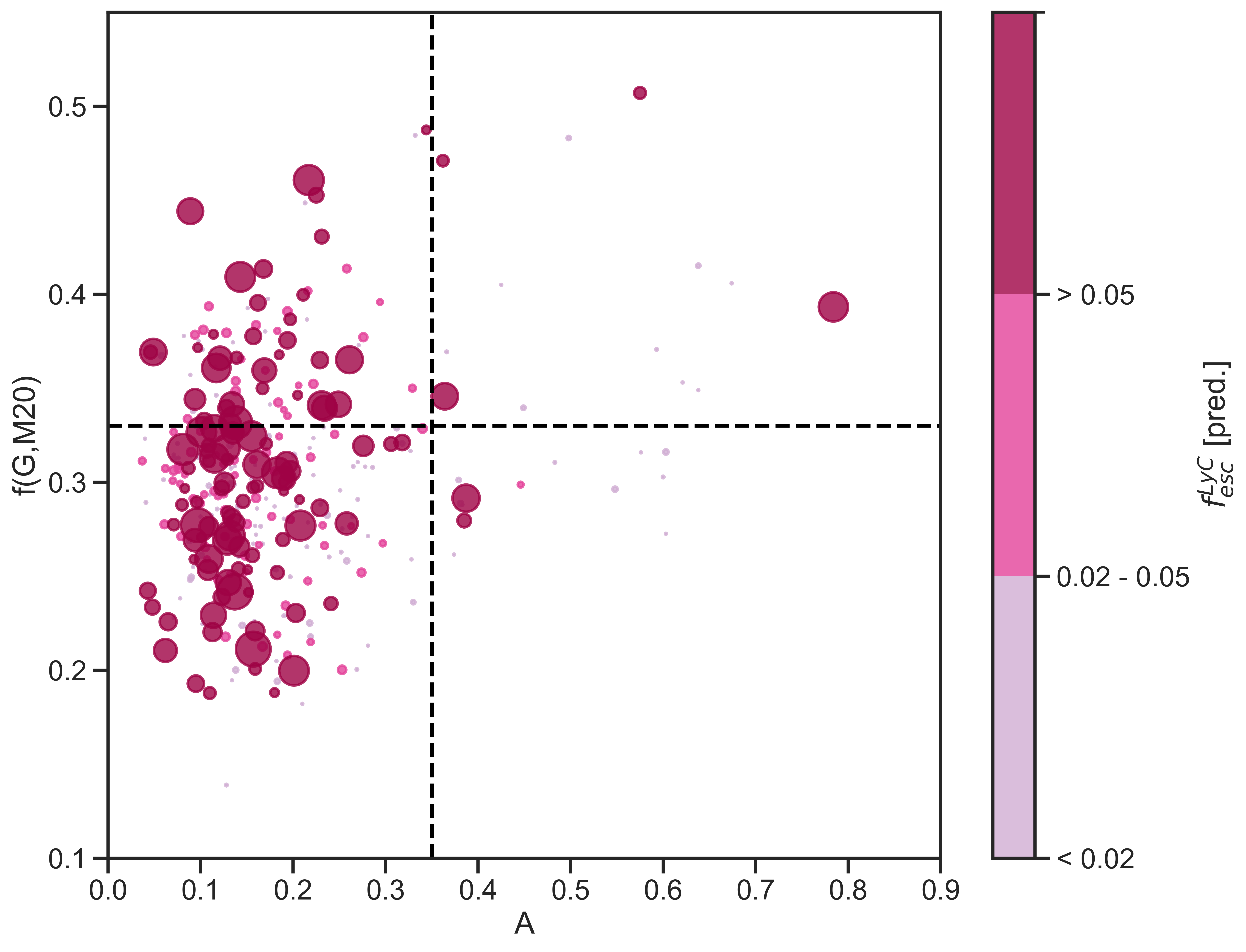}
\caption{Galaxies in our sample, color-coded according to their predicted $f_{\rm esc}$ values in three ranges: $f_{\rm esc} < 0.02$ (pale violet), $0.02 \leq f_{\rm esc} \leq 0.05$ (magenta), and $f_{\rm esc} > 0.05$ (dark red). The point size reflects the predicted 
$f_{\rm esc}$, with larger points indicating higher values. \label{fig:fesc_morp}}
\end{figure}

\begin{figure}[ht!]
\centering
\includegraphics[width=0.9\linewidth]{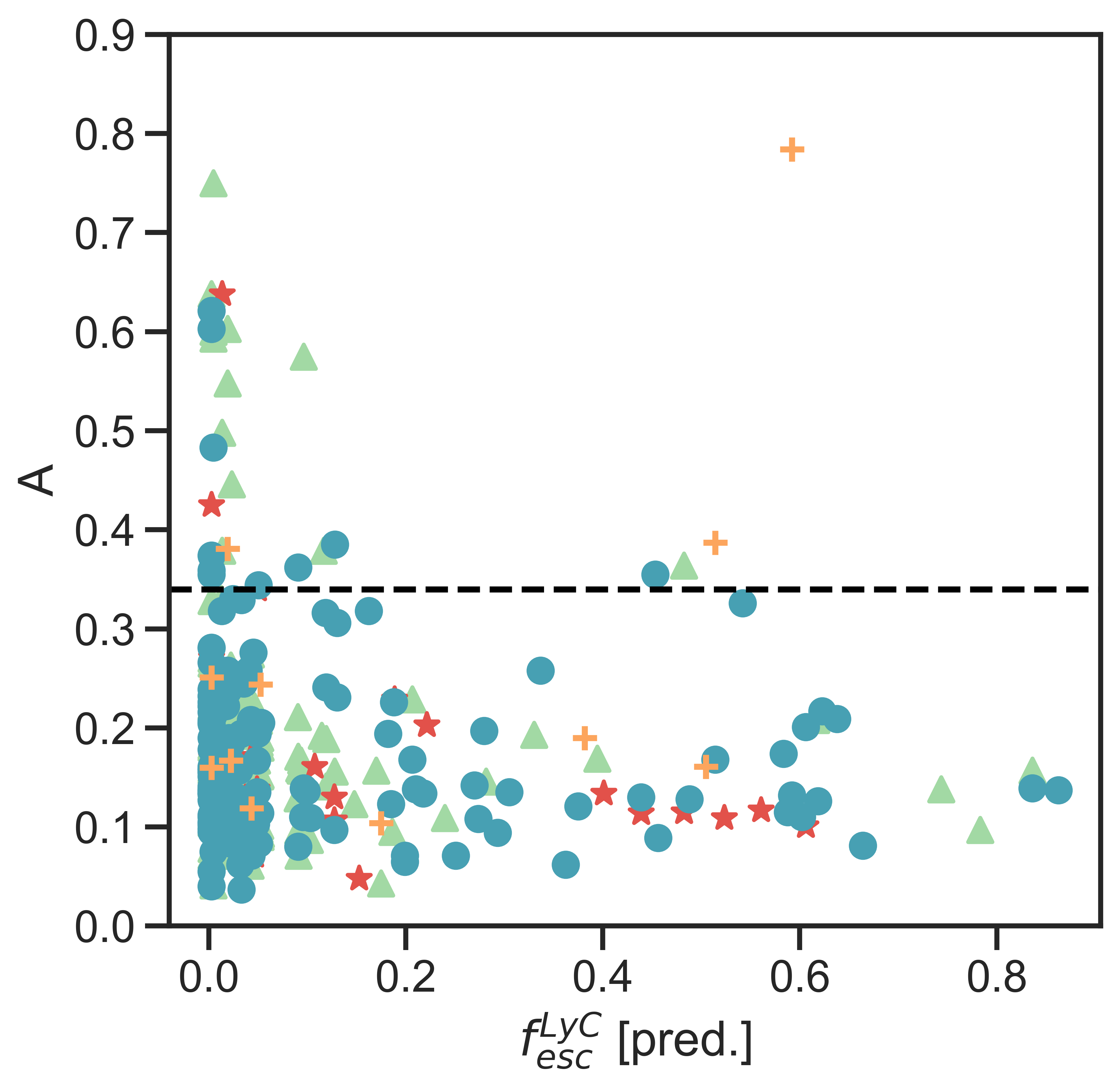}
\caption{Predicted $f_{\rm esc}$ as a function of the $A$ parameter.} \label{fig:fesc_A}
\end{figure}

 We examine the presence of merger and their connection to the predicted LyC escape fraction $f_{\rm esc}$ for our sample of galaxies at $z = 5-7$. In Fig.~\ref{fig:fesc_morp} we color-coded the sources by the predicted $f_{\rm esc}$ and we found that high $f_{\rm esc}$ sources are equally distributed between the mergers and non mergers classification. Specifically, we focus on $f_{\rm esc}$ values exceeding 0.05, which is considered a reliable threshold for identifying strong LCEs, as suggested by \cite{Flury2022}. This choice is supported by our previous works \citep{Mascia2023_Glass, Mascia2024_CEERS}, which demonstrated that although there are uncertainties associated with these $f_{\rm esc}$ values, those within this range are more reliable. Specifically, 4 of the 13 mergers in the Gold sample ($31 \pm 13$\%), 34 of 78 mergers in the Silver sample ($44 \pm 6$\%) and 107 of the 329 non-mergers ($31 \pm 3$\%) exhibit high LyC leakage.  The fractions are thus all consistent across all groups, even if the larger uncertainties in some subsets, particularly those with lower populations, prevent us from drawing firm conclusions. Note that the conclusions remain unchanged even if we consider 0.1 as a threshold. In Fig.~\ref{fig:fesc_A}, we observe a potential connection between galaxy asymmetry ($A$) and $f_{\rm esc}$, with a Pearson correlation coefficient of $r = -0.08$ and a corresponding $p$-value of 0.17. In particular, lower $f_{\rm esc}$ values are associated with the full range of $A$ values, whereas higher $f_{\rm esc}$ values appear to be exclusively linked to galaxies with low asymmetry. It is important to remember that the predictions for $f_{\rm esc}$ in the two models we are using are independent of any morphological terms, in contrast to other models previously employed \citep{Mascia2023_Glass, Mascia2024_CEERS, Jaskot24a}. Our results do not point to an increased fraction of high leakers in the merging systems, but rather to the fact that the strongest leakers are preferentially compact. 

Recently, \cite{Bhagwat2023} investigated the link between morphology and LyC escape in the \textsc{spice} simulations: their results 
support the idea that internal processes, rather than interactions, play a more critical role in driving LyC escape. Their simulations show that bursty stellar feedback can create low-density channels within galaxies that allow LyC photons to escape, regardless of whether the galaxy is undergoing a merger. This feedback-driven morphology, especially in dispersion-dominated systems, enhances LyC escape by producing irregular structures through internal processes rather than external interactions. This picture is consistent with \cite{Flury2024} who presented direct evidence for bursty star formation as a mechanism that enhances $f_{\rm esc}$ in LCEs compared to non-LCEs.

These findings align with our results, which show no relation between mergers and the predicted LyC escape, and support the view that LyC escape at these redshifts is predominantly governed by intrinsic factors such as the compactness of the star formation region.
This observation is consistent with results from the LzLCS+ survey, where \cite{Flury2022} and \cite{Jaskot24a} identified compactness as a strong predictor of $f_{\rm esc}$, emphasizing its role as a key driver of the strong dependence of $f_{\rm esc}$ on $\Sigma_{\text{SFR}}$. Moreover, in a study of a leaker at $z = 3.088$, \cite{Gupta2024} discussed how optically thin channels, potentially created by merger-driven outflows, enable LyC escape. However, they caution that the complex outflow dynamics typical of high-redshift systems limit the likelihood of star-formation-driven feedback alone creating such channels.

Despite the strengths of our morphological classification, some limitations must be acknowledged. First, our classification is based on rest frame UV imaging, which may not adequately capture the presence of merger signatures at $z \sim 7$. Applying our classification to the optical rest-frame morphology might have resulted in a different outcome, although \cite{Treu2023} found that the morphologies of the galaxy remain consistent across wavelengths from the optical rest-frame to UV.
Another limitation arises from the uncertainty in classifying merger versus non-merger sources  based solely on parametric classification. While we conducted careful visual inspections for the sources in the Gold and Silver samples, this method may introduce bias and overlook small morphological features indicative of ongoing interactions or mergers. 

As a final remark, it is important to consider the timing of mergers: what we identify in our sample could represent pre-merger conditions, or early stages of mergers, rather than fully merging systems. The formation of low density channels that facilitate LyC escape may occur at later stages of the mergers, e.g. specifically during the coalescence phase \citep[e.g., ][]{Conselice_2000, Lotz_2004, Pawlik2016} which are not captured by our classification scheme.


\subsection{\texorpdfstring{Merger fraction versus $\xi_{\rm ion}$}{Merger fraction versus xi ion}}

We also investigate the relationship between mergers and the ionizing photon production efficiency $\xi_{\rm ion}$. 
Mergers are typically thought to temporarily increase star formation rates (SFR) by compressing gas and triggering starbursts, which in turn could increase the production of ionizing photons \citep[e.g.,][]{Barnes1991, Mihos1996}. Higher SFR typically corresponds to a younger, massive stellar population, which emits more ionizing photons due to its high luminosity in both the UV and ionizing continua. A strong correlation between specific star formation rate (sSFR) and $\xi_{\rm ion}$ has been observed \citep{Castellano2023}, making sSFR one of the most reliable tracers of a galaxy's ionizing efficiency, especially in the context of cosmic reionization \citep[e.g.,][Llerena et al. subm.]{Castellano2023}. \cite{Witten2024}, studying a sample of \lya\ emitters at $z>7$, further support this connection, demonstrating that mergers can trigger episodic bursts of star formation, which create low-density channels in the ISM, facilitating the escape of both \lya\ and ionizing photons, and potentially enhancing $\xi_{\rm ion}$ during merging events. However, compact low-mass galaxies with relatively high SFRs have also been shown to exhibit elevated $\xi_{\rm ion}$ \citep{Castellano2023}, regardless of whether they are undergoing mergers.

\begin{figure}[ht!]
\centering
\includegraphics[width=\linewidth]{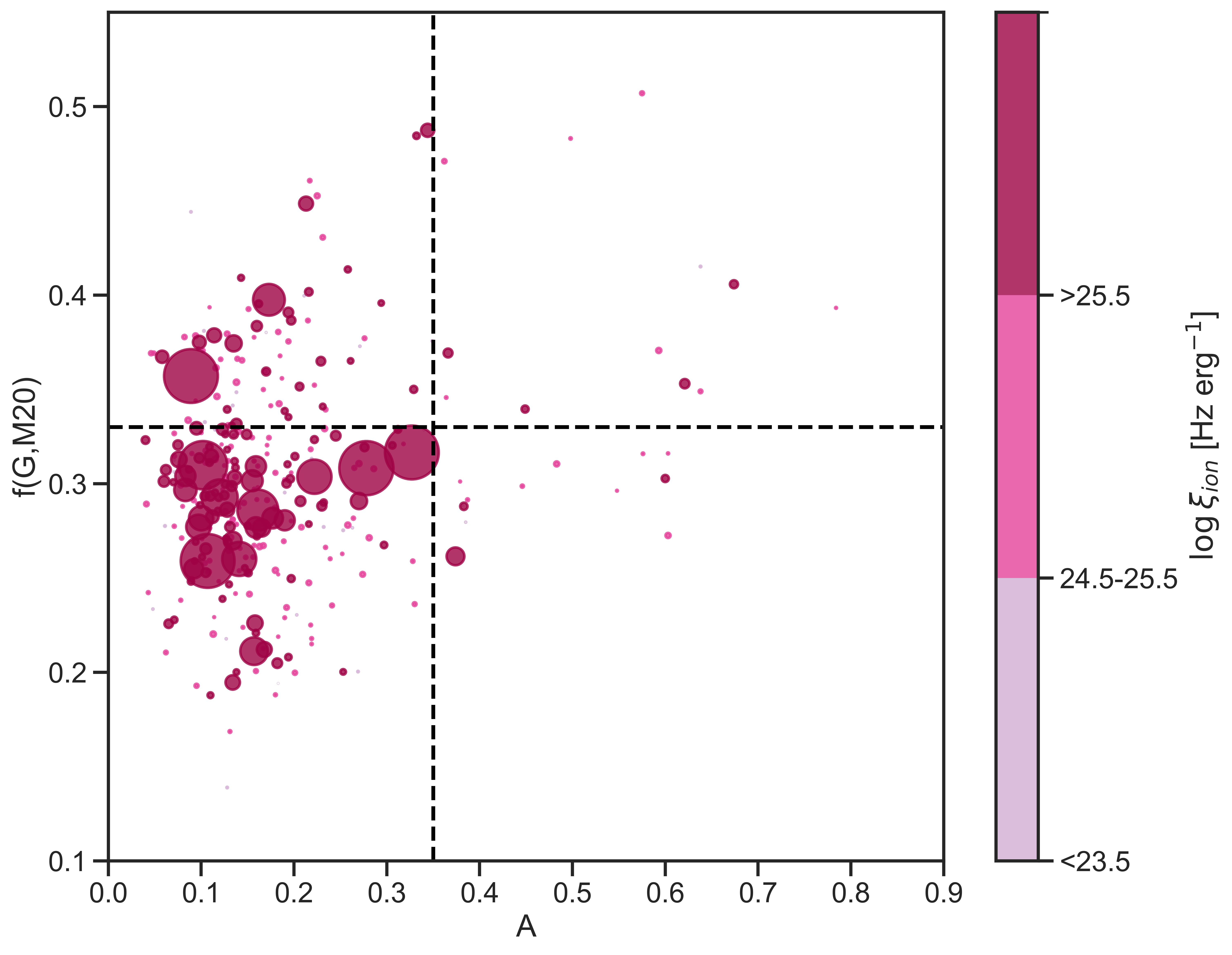}
\caption{Morphological parameters for galaxy in our sample, color-coded according to their measured $\log \xi_{\rm ion}$ values in three ranges: $23.5 \leq \log\xi_{\rm ion} \leq 24.5$ (pale violet), $24.5 \leq \log\xi_{\rm ion} \leq 25.5$ (magenta), and $\log\xi_{\rm ion} > 25.5$ (dark red). The point size reflects the $\log\xi_{\rm ion}$, with larger points indicating higher values.\label{fig:xiion_morp}}
\end{figure}

Our results are presented in Fig.~\ref{fig:xiion_morp}, where we plot the morphological parameters of our galaxies and color code them according to their $\xi_{\rm ion}$. We can see how galaxies with elevated $\xi_{\rm ion}$ are distributed across both merging and non-merging systems. Applying a threshold of $\log\xi_{\rm ion} = 25.5$, we find that 4 of the 13 mergers in the Gold sample ($31 \pm 13$\%), 28 of 78 mergers in the Silver sample ($36 \pm 6$\%), and 133 of the 329 non-mergers ($39 \pm 3$\%) have elevated ionizing emissivity. 
Although errors are large for samples exhibiting merger characteristics, the results suggest that elevated values of $\log\xi_{\rm ion}$ might be more common in compact, symmetric sources. These conclusions remain unchanged even if we consider 25.6 as a threshold.

Additionally, the few sources with very high  $\log\xi_{\rm ion}$ all  exhibit very compact and symmetric morphologies in the rest-frame UV, reinforcing the idea that the production and escape of LyC radiation during the EoR are more closely related to compactness than to galaxy interactions or mergers.

\section{Summary and Conclusions}\label{sec:conclusions}

In this work, we assembled a sample of 436 spectroscopically confirmed sources at redshifts $5 \leq z \leq 7$ from various JWST surveys (EIGER, CEERS, DD-2750 and JADES). By performing SED fitting with  \texttt{Prospector} \citep{Johnson2021}, we derived their physical properties such as stellar mass ($M_*$), UV magnitude ($M_{\rm UV}$), UV $\beta$ slope, dust reddening (E(B-V)), and specific star formation rate (sSFR). Emission line measurements allowed us to determine $\log\xi_{\rm ion}$ and other key properties, such as the O32 emission line ratio and the EW(H$\beta$) and EW([O\textsc{iii}]). Finally, using a well-tested morphological scheme, we classified the morphology of all systems according to \cite{Dalmasso2024}, which employ a combination of Asymmetry (A), Gini (G) and $M_{20}$ parameters.

We then compared the properties of these high-redshift sources to the LzLCS+ sample, and we found that low- and intermediate-redshift LyC emitters are not always analogs for cosmic reionizers, showing a diverse range of properties that only partially overlap with galaxies in the EoR. For this reason,  we identified a subsample of the LzLCS+ survey sources, comprising 51 galaxies out of the original 88, that are the best analogs for the sources we are characterizing during the EoR.

Although we have restricted the LzLCS+ sample and created the subsample of ``best analogs'', it is important to realize that a proof that the latter is truly representative of the LyC emissivity of EoR galaxies and its correlation with other properties of galaxies does not yet exist. The gaseous environment, in particular the ISM and the CGM, of low-z galaxies is most likely profoundly different from that of galaxies at the EoR. For example, already at cosmic noon, i.e. $z \sim 2-3$, as reviewed by \cite{Tacconi2018}, the gas fraction of star-forming disks is much larger, up to 80\%, than in the local Universe and this can potentially have profound consequences on $f_{\rm esc}$.  Figure \ref{fig:comparison_models} shows good agreement between the observed properties of galaxies at $z \sim 3$ and those inferred from the Cox models based on the LzLCS+, which offers some degree of reassurance that our sample of ``best analogs'' captures the key features of the correlation between $f_{\rm esc}$ and galaxy properties: however, it is important to keep in mind that a final verification is still to be made.

Assuming that the analogy holds, using this subsample of analogs, we propose two refined Cox models to predict $f_{\rm esc}$ during the EoR. These models are based on $M_{\rm UV}$, E(B-V), $\log_{10}(M_*)$, and $\log_{10}(\text{EW}([\text{O}\textsc{iii}]+\text{H}\beta))$ or $\log_{10}(\text{O32})$.The new models perform as well as, and in one case better than, the models originally proposed by \cite{Jaskot24b}, as demonstrated by the concordance index ($C$).
To further probe the robustness of our new models, we applied them to a small sample of confirmed LyC leakers at $z\approx3$ (the only ones where all the necessary properties can be measured) and found that the models successfully predict the escape fractions of these sources.

We applied the revised Cox models to infer $f_{\rm esc}$ for our EoR galaxies. Combining these predictions to the identification of merging systems, we conducted an analysis of the correlation between the presence of merger features and the production and escape of ionizing photons in galaxies during the EoR.
Our main findings are as follows:
\begin{itemize}
\item When we applied the new models to the large sample of galaxies observed during the EoR, we confirmed that the majority exhibit modest escape fractions with median values of $f_{\rm esc} \sim 0.04$, and an average average $f_{\rm esc} \sim 0.13$. The distribution of predicted escape fractions is highly skewed with most of the objects showing small escape fractions but with a tail extending to very high values. 
This finding reinforces previous conclusions by \cite{Mascia2024_CEERS}, suggesting that the galaxies we are currently probing (with $M_{\rm UV}$ as faint as $-18$) contributed less than one-third to the overall reionization budget. Further exploration of the  fainter, low-mass galaxy population at high redshift is essential to fully understand their role in cosmic reionization.
\item Our analysis reveals no correlation between the predicted $f_{\rm esc}$ and the presence of merger signatures, with most galaxies in the sample exhibiting compact, symmetric morphologies rather than disturbed or irregular structures typically associated with mergers. 
\item We find an increase of $\log \xi_{\rm ion}$ at fainter $M_{\rm UV}$, which aligns with recent studies in literature, both spectroscopic and photometric \citep[e.g.,][]{Simmonds2024, Prieto-Lyon2023, Llerena2024}. We do not find indication for an increased fraction of systems with high photon production efficiency within mergers. In contrast, there are indications that the most compact systems might exhibit a higher $\xi_{\rm ion}$, suggesting a potential link between compactness and ionizing photon production.
\end{itemize}

Unfortunately, there are yet no systematic studies on the galaxy morphology of LyC leakers at low and intermediate redshifts, where direct detection of ionizing flux is feasible. Current samples at low redshifts (e.g., the LzLCS+) might also suffer from biases, as many sources were traditionally selected based on their compactness, which was thought to be related to the escape of LyC photons \cite{Izotov2018a}. As described in the Introduction, high LyC escape fractions have been reported for some merging systems at $z=1$ \citep{Maulick2024} and $z=3$ \citep{Gupta2024,Yuan2024}. Recently, \cite{zhu2024} claimed that at $z=3$, the majority of LyC leakers exhibit merging signatures, a result that would be in stark contrast to our findings in the EoR. However, their sample includes only a few confirmed leakers, as many of the sources analyzed in their work are tentative detections. The few, solid LyC leakers at $z=3$, such as Ion2, exhibit very compact morphology with no indication of any merging activity.
Additionally, their merging classification differs from the one employed in our analysis, which could affect the outcome.
We must also consider that at $z \sim 3$, the transmission of the IGM becomes highly stochastic, allowing for the detection of only the strongest LyC leakers.


This highlights the importance of building a statistically larger sample of leakers at $z \sim 3$, with a broad range of $f_{\rm esc}$ measurements (or upper limits). One example is the Parallel Ionizing Emissivity Survey (HST program 17147, PI C. Scarlata), conducted with HST to identify new LyC leakers at $3.1<z<3.5$ among a sample of $\sim 700$ galaxies. An alternative approach will be employed by the LyC22 JWST program (GO 1869, PI D. Schaerer), which has obtained NIRpec observations of known LCE. This program will provide the properties of the first robust sample of LCEs at the highest redshift where LyC can still be directly detected, offering a unique opportunity to also study their morphological properties in a systematic way. Crucially, the same data set will also allow us to recalibrate the Cox model at $z \sim 3$.  By bridging the gap between $z = 0.3$ and the EoR, this program will significantly enhance our understanding of the role of mergers and other processes in driving LyC escape, thereby improving the predictive power of $f_{\rm esc}$ during reionization.

\begin{acknowledgements}
Funded by the European Union (ERC, AGENTS,  101076224). Views and opinions expressed are however those of the author(s) only and do not necessarily reflect those of the European Union or the European Research Council. Neither the European Union nor the granting authority can be held responsible for them. 
We acknowledge support from the INAF Large Grant 2022 ``Extragalactic Surveys with JWST'' (PI Pentericci). 
We acknowledge support from INAF Mini-grant ``Reionization and Fundamental Cosmology with High-Redshift Galaxies'' and from PRIN 2022 MUR project 2022CB3PJ3 - First Light And Galaxy aSsembly (FLAGS) funded by the European Union – Next Generation EU. 
RA acknowledges financial support from  project PID2023-147386NB-I00 and the Severo Ochoa grant CEX2021-001131-S funded by MCIN/AEI/10.13039/50110001103. The project that gave rise to these results received the support of a fellowship from the “la Caixa” Foundation (ID 100010434). The fellowship code is LCF/BQ/PR24/12050015. LC acknowledges support from grants PID2022-139567NB-I00 and PIB2021-127718NB-I00 funded by the Spanish Ministry of Science and Innovation/State Agency of Research  MCIN/AEI/10.13039/501100011033 and by “ERDF A way of making Europe”

\end{acknowledgements}
\bibliographystyle{aa}
\bibliography{bib.bib} 

\newpage
\section*{Appendix A: revised Cox models} \label{appendixA}

The LzLCS+ results indicate that $f_{\rm esc}$ correlates with various galaxy properties, including line-of-sight factors like HI covering fraction, dust attenuation, and \lya\ escape fraction, as well as global properties such as O32 ratio and star formation rate surface density $\Sigma_{\text{SFR}}$ \citep[e.g.,][]{Flury2022b, Chisholm2022, Saldana-Lopez2022, Xu2023}. These correlations suggest that mechanical or radiative feedback may help create low optical depth sight lines that allow ionizing photons to escape. onetheless, the observed relationships between $f_{\rm esc}$ and various galaxy properties show considerable scatter \citep{Wang2021, Flury2022b, Chisholm2022, Saldana-Lopez2022, Xu2023}. This suggests that no single property can reliably predict $f_{\rm esc}$ on its own, and a more accurate approach likely involves combining multiple factors.

Traditional methods like linear regression often fall short in handling censored data, which contain upper limits rather than precise measurements \citep{Mascia2023_Glass, Lin2023}. To overcome these limitations, \cite{Jaskot24a, Jaskot24b} applied survival analysis techniques, particularly the Cox proportional hazards model \citep{Cox1972}, which is well-suited for datasets with censored data.

In the Cox Model, the hazard function is expressed as:

\begin{equation}
h(t | x) = h_0(t) \exp \left( \sum_{i=1}^{n} \beta_i (x_i - \bar{x}_i) \right),
\end{equation}
\noindent
where $h(t | x)$ is the hazard of detecting LyC, $h_0(t)$ is the baseline hazard function, and $\beta_i$ are the coefficients for the galaxy properties $x_i$. 

Using the absorbed fraction of LyC ($f_{\text{abs}} = 1 - f_{\rm esc}$), the survival function $S(f_{\text{abs}})$, representing the probability that $f_{\text{abs}}$ exceeds a given threshold, is:

\begin{equation}
S(f_{\text{abs}}) = \exp\left[ -H_{F_0}(f_{\text{abs}}) \cdot \text{ph}(x) \right],
\end{equation}
\noindent
where $H_{F_0}(f_{\text{abs}})$ is the baseline cumulative hazard function:

\begin{equation}
H_{F_0}(f_{\text{abs}}) = \int_0^{f_{\text{abs}}} h_0(f) \, df,
\end{equation}
\noindent
and $\text{ph}(x)$ is the partial hazard function:

\begin{equation}
\text{ph}(x) = \exp \left( \sum_{i=1}^{n} \beta_i (x_i - \bar{x}_i) \right).
\end{equation}

The best-fit coefficients $\beta_i$ in the Cox proportional hazards model are determined by maximizing the partial likelihood, which compares $\text{ph}(x)$ for each detection with the sum of $\text{ph}(x)$ for all galaxies with higher $f_{\text{abs}}$ values \citep{Breslow1972}. The baseline cumulative hazard function $H_{F_0}(f_{\text{abs}})$ is derived using Breslow’s estimator \citep{Breslow1972}, accounting for both detections and non-detections.

The median $f_{\text{abs}}$ is found when the survival function $S(f_{\text{abs}})$ reaches 0.5, giving a 50\% probability that the true $f_{\text{abs}}$ is above or below this value. Thus, the median predicted $f_{\rm esc}$ is calculated as $1 - \text{median } f_{\text{abs}}$. The survival function also provides $1\sigma$ confidence intervals (where $S(f_{\text{abs}})$ is 0.159 and 0.841), reflecting uncertainties in $f_{\rm esc}$ due to observational scatter and intrinsic variability in the galaxy population.

The Cox model’s performance is evaluated using the concordance index ($C$) \citep{Harrell1982}, which ranges from 0 (perfect disagreement) to 1 (perfect concordance):

\begin{equation}
C = \frac{n_c + 0.5n_t}{n_c + n_d + n_t},
\end{equation}
\noindent
where $n_c$, $n_d$, and $n_t$ represent concordant, discordant, and tied pairs of observed and predicted values, respectively.

In Tables~A.1 and~A.2, we provide the information necessary to predict $f_{\rm esc}$\ using the models discussed in this paper.

\begin{table}[ht!]
\caption*{Table A.1: Revised ELG-EW Model} 
\label{tab:ELG_EW}
\begin{tabular}{llll}
\hline
\multicolumn{4}{c}{{$C = 0.79$}} \\
\hline
\multicolumn{4}{c}{{Model Parameters}} \\
\multicolumn{2}{l}{Variable} & $b_i$ & ${\bar x_i}$ \\
\hline
\multicolumn{2}{l}{$M_{\rm 1500}$} & -0.69 & -19.649 \\
\multicolumn{2}{l}{$\log_{10}$($M_*$/$M_\odot$)} & -0.31 & 8.559 \\
\multicolumn{2}{l}{E(B-V)$_{\rm UV}$} & -13.25 & 0.103 \\
\multicolumn{2}{l}{$\log_{10}$(EW([\textrm{O}\textsc{iii}]+H$\beta$)/\AA)} & 3.28 & 2.941 \\
\hline
\multicolumn{4}{c}{{Baseline Cumulative Hazard}} \\
$f_{\text{abs}}$ & HF$_0$($f_{\text{abs}}$) & $f_{\text{abs}}$ (cont.) & HF$_0$ (cont.) \\
\hline
0.111 & 0.011825 & 0.9732 & 0.72714 \\
0.3753 & 0.025212 & 0.9736 & 0.790193\\
0.4162 & 0.039181 & 0.9752 & 0.790193\\
0.5089 & 0.053583 & 0.9769 & 0.855688\\
0.5667 & 0.069132 &0.978 & 0.926178\\
0.6948 & 0.085736 &0.9795 & 0.926178\\
0.7341 & 0.102827 &0.9802 & 1.004294\\
0.8079 & 0.120801 &0.981 & 1.004294\\
0.8223 & 0.141086 &0.9812 & 1.110345\\
0.8719 & 0.16315 &0.9837 & 1.221632\\
0.8803 & 0.185655 &0.984 & 1.221632\\
0.8815 & 0.209659 &0.9849 & 1.221632\\
0.8947 & 0.234525 &0.986 & 1.221632\\
0.9083 & 0.260956 &0.9867 & 1.221632\\
0.9102 & 0.288869 &0.9868 & 1.444606\\
0.9411 & 0.288869 &0.9876 & 1.688703\\
0.9471 & 0.288869 &0.9886 & 1.688703\\
0.9481 & 0.322321 &0.9893 & 1.688703\\
0.9507 & 0.356882 &0.9906 & 1.688703\\
0.9527 & 0.392632 &0.9911 & 1.688703\\
0.9569 & 0.430956 &0.9931 & 1.688703\\
0.9579 & 0.472974 &0.9955 & 1.688703\\
0.9624 & 0.517262 &0.9956 & 3.460356\\
0.9667 & 0.564117 &0.9975 & 3.460356\\
0.9691 & 0.614575 & \\
0.9693 & 0.668437 & \\
0.9709 & 0.668437 & \\
\hline
\end{tabular}
\end{table}

\begin{table}[ht!]
\centering
\caption*{Table A.2: Revised ELG-O32 Model} \label{tab:ELG_O32}
\begin{tabular}{llll}
\hline
\multicolumn{4}{c}{{$C = 0.83$}} \\
\hline
\multicolumn{4}{c}{{Model Parameters}} \\
\multicolumn{2}{l}{Variable} & $b_i$ & ${\bar x_i}$ \\
\hline
\multicolumn{2}{l}{$M_{\rm 1500}$} & -1.63 & -19.649 \\
\multicolumn{2}{l}{$\log_{10}$($M_*$/$M_\odot$)} & -0.76 & 8.559 \\
\multicolumn{2}{l}{E(B-V)$_{\rm UV}$} & -13.87 & 0.104 \\
\multicolumn{2}{l}{$\log_{10}$(O32)} & 6.55 & 0.843 \\
\hline
\multicolumn{4}{c}{{Baseline Cumulative Hazard}} \\
$f_{\text{abs}}$ & HF$_0$($f_{\text{abs}}$) & $f_{\text{abs}}$ (cont.) & HF$_0$ (cont.) \\
\hline
0.111 & 0.003188 & 0.9732 & 0.732829 \\
0.3753 & 0.009485 &0.9736 & 0.832722 \\
0.4162 & 0.016245 &0.9752 & 0.832722 \\
0.5089 & 0.024456 &0.9769 & 0.939902 \\
0.5667 & 0.033316 &0.978 & 1.057705 \\
0.6948 & 0.042431 &0.9795 & 1.057705 \\
0.7341 & 0.051797 &0.9802 & 1.188196 \\
0.8079 & 0.061708 &0.981 & 1.188196 \\
0.8223 & 0.0772 &0.9812 & 1.381873 \\
0.8719 & 0.096801 &0.9837 & 1.590329 \\
0.8803 & 0.116622 &0.984 & 1.590329 \\
0.8815 & 0.137886 &0.9849 & 1.590329 \\
0.8947 & 0.159401 &0.986 & 1.590329 \\
0.9083 & 0.182399 &0.9867 & 1.590329 \\
0.9102 & 0.206052 &0.9868 & 1.913137 \\
0.9411 & 0.206052 &0.9876 & 2.32817 \\
0.9471 & 0.206052 &0.9886 & 2.32817 \\
0.9481 & 0.238704 &0.9893 & 2.32817 \\
0.9507 & 0.273793 &0.9906 & 2.32817 \\
0.9527 & 0.315661 &0.9911 & 2.32817 \\
0.9569 & 0.359126 &0.9931 & 2.32817 \\
0.9579 & 0.406118 &0.9955 & 2.32817 \\
0.9624 & 0.455871 &0.9956 & 4.960163 \\
0.9667 & 0.51674 &0.9975 & 4.960163 \\
0.9691 & 0.58189 & \\
0.9693 & 0.648165 &\\
0.9709 & 0.648165 &\\
\hline
\end{tabular}
\end{table}
\end{document}